\documentclass[useAMS,usenatbib]{mn2e}

\usepackage{verbatim}
\usepackage{graphicx}
\usepackage{multirow}
\usepackage{subfigure}

\title[The Soft X-ray Lightcurves of Partially Eclipsed Stellar Flares]{The Soft X-ray Lightcurves of Partially Eclipsed Stellar Flares}

\author[Johnstone et al.]{C. P. Johnstone$^1$, S. G. Gregory$^{2,3}$, M. M. Jardine$^1$, K. V. Getman$^4$\\$^1$ School of Physics and Astronomy, University of St Andrews, St Andrews, Fife, KY16 9SS, UK \\$^2$ School of Physics, University of Exeter, Stocker Road, Exeter, EX4 4QL, UK \\$^3$ California Institute of Technology, MC 249-17, Pasadena, CA 91125, U.S.A. \\$^4$ Department of Astronomy and Astrophysics, 525 Davey Laboratory, Pennsylvania State University, University Park, PA 16802, USA}

\begin{document}

\maketitle

\begin{abstract}
Most stellar flares' soft X-ray lightcurves possess a `typical' morphology, which consists of a rapid rise followed by a slow exponential decay.
However, a study of 216 of the brightest flares on 161 pre-main sequence stars, observed during the \emph{Chandra} Orion-Ultradeep Project (COUP), showed that many flare lightcurves depart from this typical morphology.
While this can be attributed to the superposition of multiple typical flares, we explore the possibility that the time-variable eclipsing of flares by their host stars may also be an important factor. 
We assume each flare is contained within a single, uniform plasma density magnetic loop and specify the intrinsic variation of the flare's emission measure with time. 
We consider rotational eclipse by the star itself, but also by circumstellar discs and flare-associated prominences. Based on this simple model, we generate a set of flares similar to those observed in the COUP database. 
Many eclipses simply reduce the flare's maximum emission measure or decay time.  
We conclude therefore that eclipses often pass undetected, but usually have only a modest influence on the flare emission measure profile and hence the derived loop lengths.  
We show that eclipsing can easily reproduce the observed atypical flare morphologies. 
The number of atypical modelled flare morphologies is however much less than that found in the COUP sample. 
The large number of observed atypical flare morphologies, therefore, must be attributed to other processes such as multiple flaring loops.
\end{abstract}

\section{Introduction} \label{sect:intro}

\begin{figure*}
\centering
\includegraphics[width=1.0\textwidth]{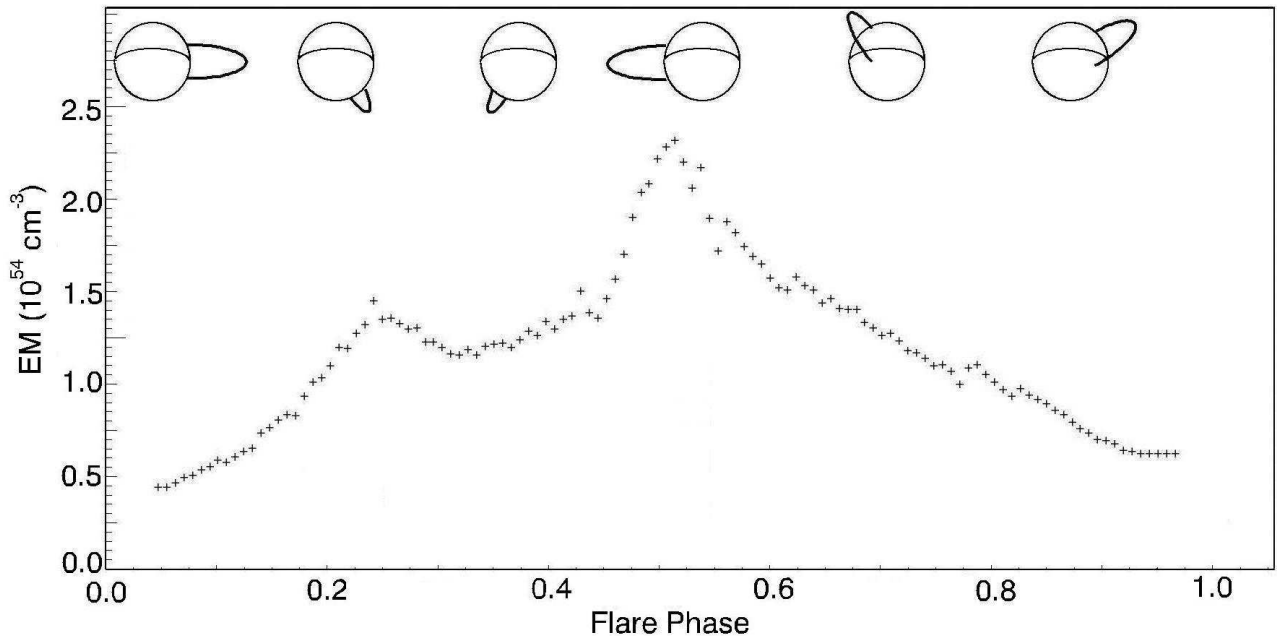}
\caption{Example flare showing how eclipsing can cause an atypical flare morphology. The flare is assumed to be contained within a single magnetic loop with a uniform plasma density. The example emission measure curve is the flare on COUP source 649 given in Fig. \ref{fig:flarefits}. The images at the top show the position of the single magnetic loop containing the flaring plasma, the geometry of which has been determined from the best fit emission measure curve given in Fig. \ref{fig:flarefits}. Flare phase 0.0 and 1.0 represent the beginning and end of the flare respectively.}
 \label{fig:example}
\end{figure*}

%

Stellar flares are generally regarded as the stellar analogues of solar flares and their X-ray emission often only differs from their solar counterparts in magnitude and duration. 
Stellar flare peak temperatures and emission measures can be orders of magnitude greater than what is seen on the Sun.
The durations of the longest lived stellar flares significantly exceed the longest durations seen in solar flares.
For a detailed comparison between solar and stellar flares, see \citet{2008ApJ...672..659A}.
They also differ in the fact that stellar flares are spatially unresolved, whereas this is only the case in the smallest of flare events on the Sun. 
Thus it is not possible to observe directly where on its host star a flare is located, or whether parts of the flare are eclipsed by the host star.
While typical durations of solar flares are very much less than the solar rotation period, this is not the case with most stellar flares.
As a result, the likelihood that a stellar flare undergoes a rotational eclipse is much greater than that for solar flares.
It is therefore natural to expect that although solar and stellar flares are probably produced by similar processes, their observational signatures may differ.

For a review of the physical mechanisms in stellar flares, we refer the reader to \citet{doi:10.1146/annurev-astro-082708-101757}.
A significant departure from a potential field configuration in a coronal magnetic field corresponds to a large amount of excess energy being held in that field.
When reconnection events occur, the coronal magnetic field geometry is simplified (i.e. becomes closer to a potential field configuration) and the resulting configuration corresponds to a lower energy state. 
In this process, a large amount of the excess energy is converted into the non-thermal motions of electrons and ions which spiral down magnetic field lines and impact the stellar chromosphere. 
This can be seen at radio wavelengths as the charged particles emit gyrosychrotron radiation.
As the energetic electrons impact the chromosphere, they emit non-thermal Bremsstrahlung radiation at hard X-ray wavelengths, as they become thermalised by random Coulomb interactions.   
This model is known as the ``thick-target model" (\citealt{1968ApJ...153L..59N}; \citealt{1971SoPh...18..489B}; \citealt{1976SoPh...50..153L}).
This causes chromospheric plasma to be heated and evaporated into the corona where it is contained within magnetic loop structures.
Through a combination of mostly radiative losses and heat conduction back to the photosphere, the evaporated plasma cools (\citealt{1978ApJ...220.1137A}).


In this paper, we consider the morphologies of the soft X-ray lightcurves of spatially unresolved stellar flares. 
In the majority of cases, the morphologies of typical flares can be broken down into two distinct phases. 
The first phase consists of a rapid increase in luminosity due to the heating and evaporation of chromospheric plasma. 
This is followed by a slow exponential decay due to cooling.
However, a large number of stellar flares show more complex atypical morphologies (\citealt{2008ApJ...688..418G}). 
Among these atypical morphologies are flares with longer rise phases and no clear peak and flares with multiple peaks or dips in their lightcurves.
The interpretation of these events is important because large flares, especially on young pre-main sequence stars, can provide information about the extent of X-ray coronae (see for example, \citealt{2006ApJS..164..173M} and \citealt{2008ApJ...688..437G}).
Ionization by large X-ray flares can significantly influence the chemistry and turbulence (via the magneto-rotational instability) of circumstellar discs, which can have profound effects on accretion, dust settling, protoplanet migration and other physical processes (\citealt{2006A&A...445..205I}, \citealt{2010HiA....15..744F})


Several interpretations of multiple peaked flares, often based on solar analogies, have been proposed.
For example, \citet{2004A&A...416..733R} observed an X-ray flare on Proxima Centauri that showed two distinct peaks in its lightcurve. 
They concluded that the second peak was probably produced through a similar event in a second loop system.  
Similarly, \citet{2010ApJ...712...78L} reported the observation, by XMM-Newton, of the unusually long ($\sim$36ks) rise phase of a flare on a young M star in the TW Hya association. 
They interpreted this rise phase as being a result of the superposition of multiple flares in separate loop systems.
This interpretation of stellar flares with similar morphologies is common in the literature (see for example \citealt{2005A&A...430..155P}, \citealt{2008MNRAS.387.1627P}).



In this work, we consider an interesting geometric alternative to the explanations given above. 
In this alternative, atypical flare morphologies are not the result of multiple flare events, but are the result of the time variable eclipsing of the flaring coronal plasma caused by the rotation of the host star.
Previous studies have used this interpretation to explain the morphologies of stellar flares using eclipsing by the flares' host stars (\citealt{1997ApJ...486..886S}, \citealt{1999A&A...344..154S}) or by a companion star in eclipsing binary systems (\citealt{1999Natur.401...44S}, \citealt{2003A&A...412..849S}, \citealt{2006A&A...445..673S}, \citealt{2007A&A...466..309S}).
In this work, we explore the eclipsing interpretation within the framework of a single loop model.
Based on the solar analogy, it has recently been argued that it is unlikely that the large stellar flares considered here take place within a single magnetic loop (\citealt{2011ApJ...730....6G}).
However, the single loop assumption is often taken as a good approximation in situations where there is a single dominant loop within a complex loop system.

The way in which eclipsing can produce flares with atypical morphologies can be seen in the following hypothetical situation, shown in Fig. \ref{fig:example}, in which a flare appears to show a double peaked morphology.
In this example, the beginning of the rise phase (flare phase equal to 0.0) occurs when the flaring magnetic loop is on the limb of the stellar disc.
Initially, as chromospheric plasma is evaporated into the corona, the visible emission measure increases. 
However, as the star rotates, flaring plasma is rotated out of view, resulting in a shallower rise from phases 0.0 to 0.25. 
As the rate at which flaring plasma is eclipsed becomes equal to and then exceeds the rate at which plasma is added to the corona, an initial peak is seen (flare phase equal to 0.23) followed by a gradual decay in the visible emission measure.
However, the host star is at an inclination angle such that the flaring loop is never totally eclipsed.  
The flare's rise phase ends at flare phase equal to 0.3 and the decay phase begins.
This, however, is not seen in the visible emission measure curve. 
As the eclipsed section of the flaring magnetic loop begins to rotate back into view, a second increase in the visible emission measure is seen.
As the rate at which the flare's total emission measure decreases, it equals and then exceeds the rate at which eclipsed plasma is rotated back into view.
Thus, a second peak followed by a second decay phase is seen in the flare's emission measure curve.


Although individual stellar flares have been studied in detail, in the last few years it has become possible to study large homogeneous samples of flares. 
The largest such study is the \emph{Chandra} Orion-Ultradeep Project (COUP; \citealt{2005ApJS..160..319G}). 
In 2003, the \emph{Chandra X-Ray Observatory} provided 13 days of near-continuous observations of the members of the Orion Nebular Cluster.
Using these observations, \citet{2005ApJS..160..319G} identified 1616 X-ray sources, of which $\sim$1400 sources were confirmed as members of the Orion star forming region and the majority of the rest being background quasars seen through the molecular cloud of the region (\citealt{2005ApJS..160..353G}).
Using the COUP data, \citet{2008ApJ...688..418G} reported the detection of 216 bright flares on 161 of these stars using the condition that a `bright flare' is any event that has a peak count rate above four times the characteristic (quiescent) count rate for the host star.
This sample consists of the longest, brightest and hottest flares detected during the COUP mission. 
\citet{2008ApJ...688..418G} derived emission measures, flare durations, flare loop lengths (using the hydrodynamic models of \citet{1997A&A...325..782R}) and other parameters which will be used extensively in this paper.  
A scheme was defined that classified flares based on their lightcurve morphologies. 
Each of the 216 flares were classified as typical (84), double (8), step (38), slow-rise top-flat (20), other (24) or incomplete (42) (see \citealt{2008ApJ...688..418G} for precise definitions and example flare morphologies).

\begin{figure*}
\centering
\subfigure[]{\includegraphics[width=0.33\textwidth]{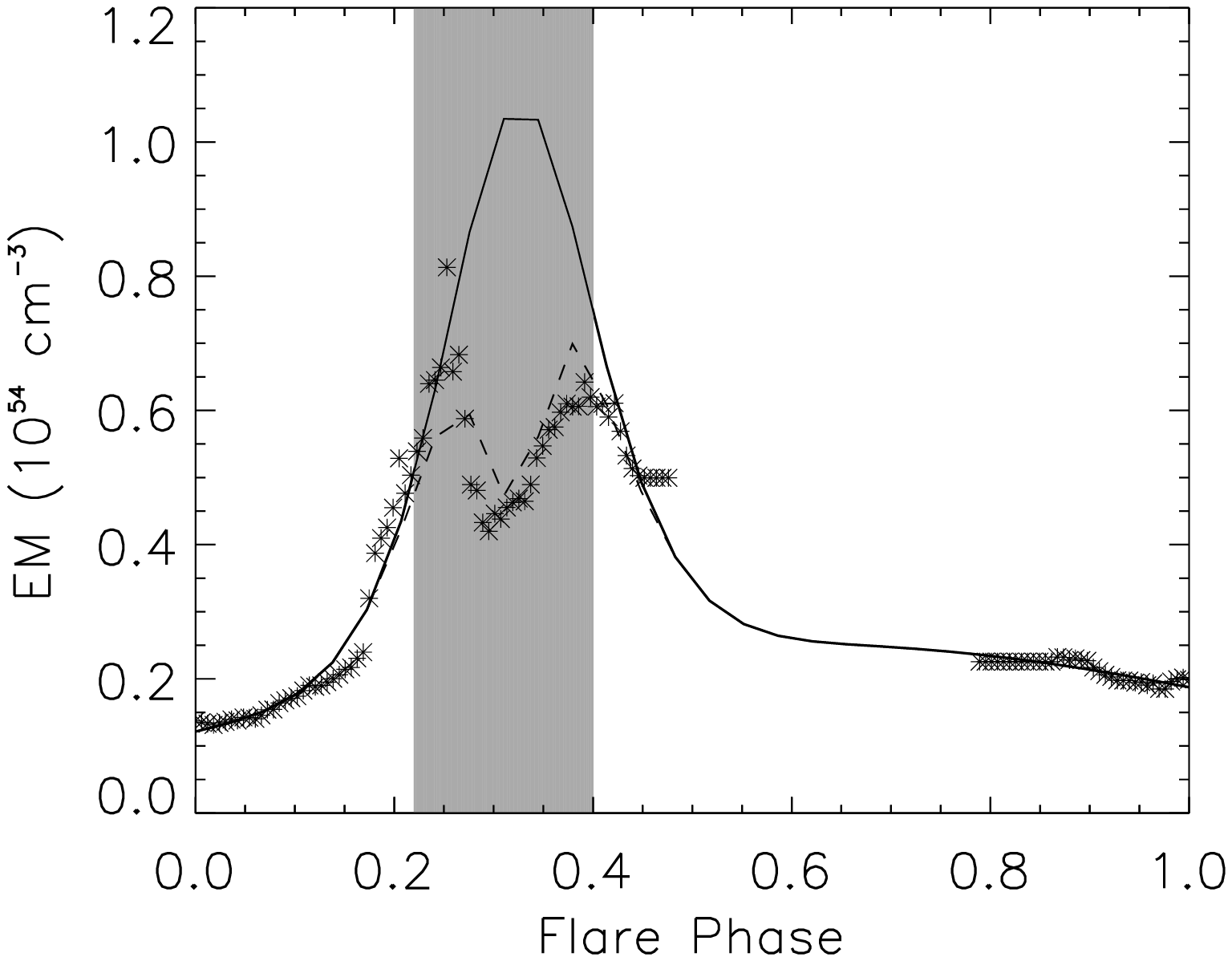}}
\subfigure[]{\includegraphics[width=0.33\textwidth]{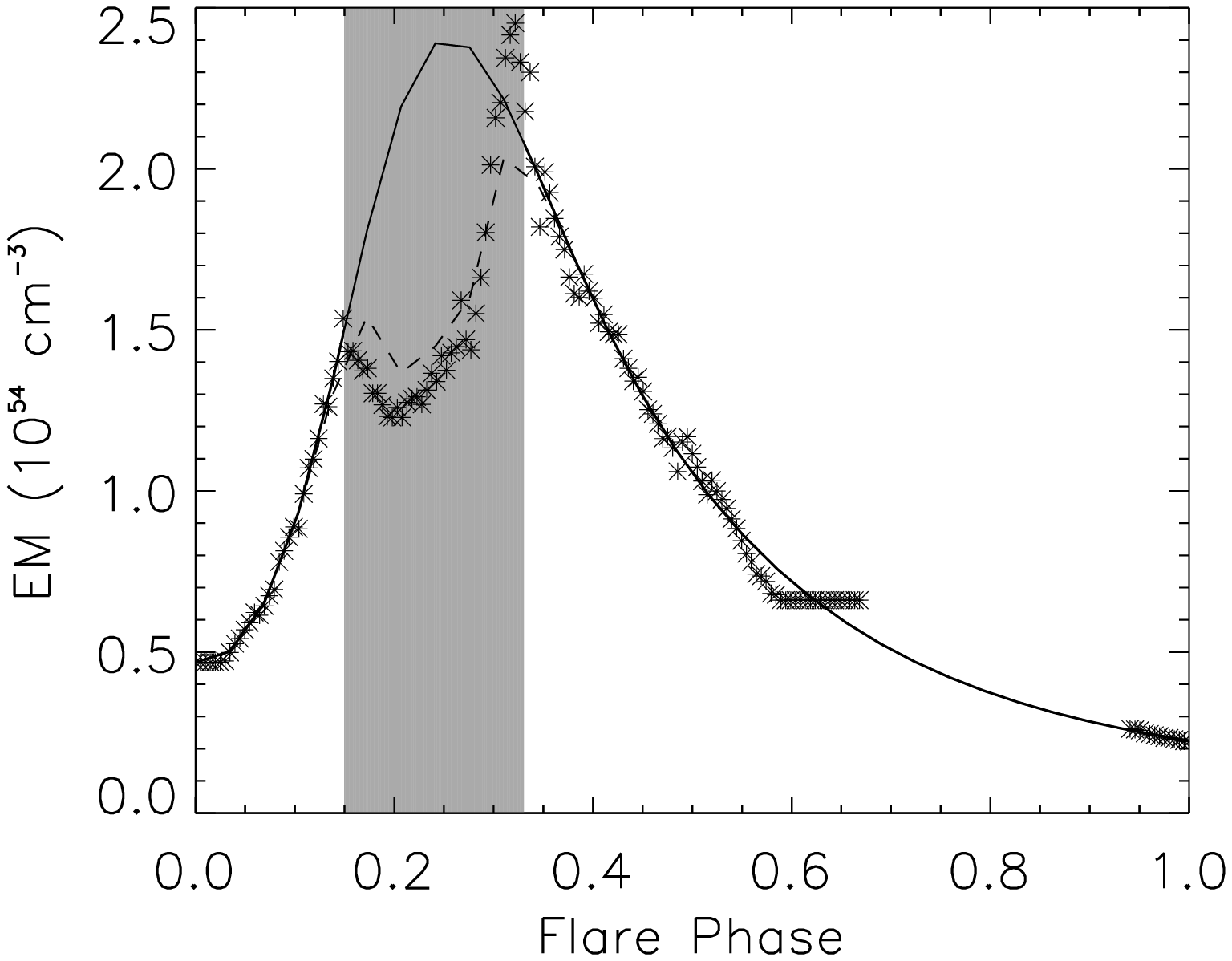}}
\subfigure[]{\includegraphics[width=0.33\textwidth]{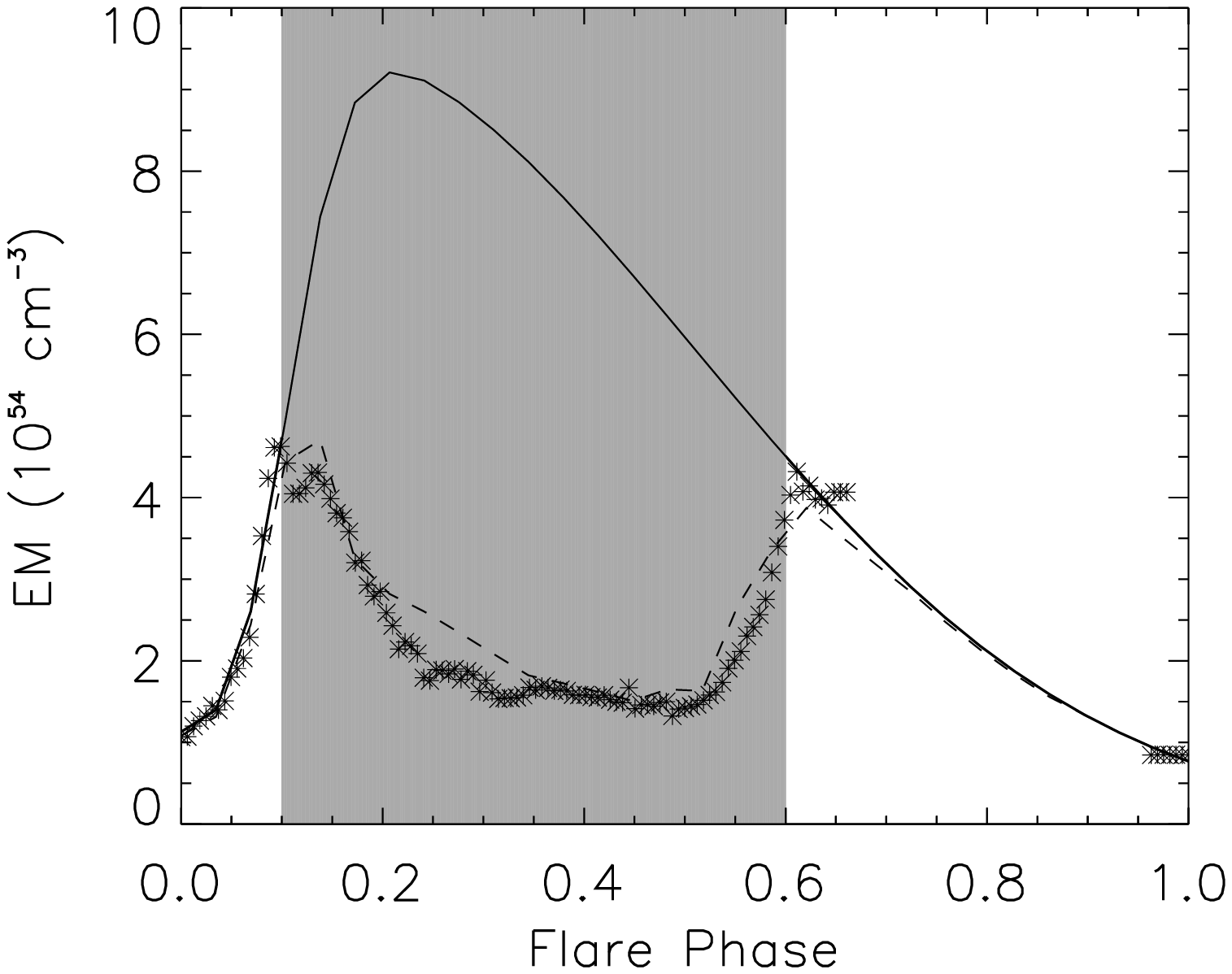}}
\caption{Emission measure vs time curves for the observed flares on COUP sources 66 (left), 649 (middle) and 942 (right) which represent good examples of atypical COUP flares. The asterisks show the observed COUP emission measure data, given by \citet{2008ApJ...688..418G}. The shaded area represents the times when some portion of the magnetic structure containing the flare was eclipsed by the host star. The solid lines show the intrinsic emission measure curves, $EM_{tot}(t)$, which have been fitted to the observed emission measure curves in the region outside the shaded area; this represents what the flare may have looked like had the flare always remained in view. The dashed line shows an eclipsed version of the same modelled flare which gives the best fit to the observed flare's emission measure curve.}
\label{fig:flarefits}
\end{figure*}

In this paper, we consider the eclipsing interpretation of atypical stellar flare morphologies.
More specifically, we ask whether such morphologies can be produced through the eclipsing of typical flares and then ask to what extent the atypical COUP flares are likely to have been produced in this way. 
In Section 2, we describe the simple flare model which is used throughout this paper. 
In Fig. \ref{fig:flarefits}, we use three examples of COUP flares to show that a range of atypical flare morphologies can be explained by eclipsing. 
In Section \ref{sect:COUPflares}, we consider the distribution of atypical COUP morphologies and compare it to a similar modelled set of flares.
In Section \ref{sect:lengths}, we consider the effect that eclipsing of flares can have on the determination of flare loop lengths.
Finally, in Section \ref{sect:summary}, the main results and our conclusions from the paper are summarised.

\begin{figure}
\centering
\includegraphics[width=0.5\textwidth]{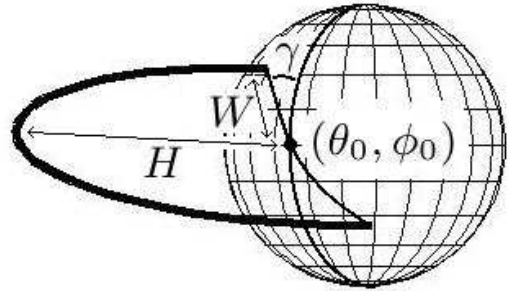}
\caption{Example of an elliptical loop used to illustrate the parameters that define a flare's geometry. $H$ is the loop height, $W$ is the loop width, $\gamma$ is the angle between the plane in which the loop is contained and the stellar rotation axis, and $\theta_0$ and $\phi_0$ are the latitude and longitude of the centre of the loop on the stellar surface.}
 \label{fig:geoparams}
\end{figure}

\section{Flare Model} \label{sect:model}

Our flare model involves the following assumptions

\begin{itemize}
\item The flare's emission is the result of a single event consisting of a rapid rise phase followed by a slow exponential decay of the flare's emission measure with time. When modelling the variation of emission measures with time, we do not consider the physical mechanisms that are responsible for triggering the flare.
\item The flaring plasma is completely contained within a single static magnetic loop with a uniform plasma density along its entire length.
\item The geometry of the flaring loop is described by an ellipse with its centre located on, and its major axis perpendicular to, the stellar surface. The thickness of the flaring loop is assumed to be a negligible fraction of its length. 
\end{itemize}

The first assumption, which is made throughout this paper, means that any deviation from a simple flare lightcurve morphology in the modelled flares can only be a result of eclipsing of the flaring plasma. 

The flaring loop geometry is thus characterized by the following five quantities: the height of the apex of the loop (i.e. the semi-major axis of the ellipse) ($H$), the width of the loop (i.e. the semi-minor axis of the ellipse) ($W$), the latitude and longitude of the centre of the ellipse ($\theta_0$, $\phi_0$), and the angle between the plane of the ellipse and the star's rotation axis ($\gamma$). 
These quantities can be seen in Fig. \ref{fig:geoparams}.
It is worth emphasising that the term `width' in this case refers to the length of the semi-minor axis of the ellipse and not the more common definition of the length between the two loop footpoints along the segment of the great circle that connects them.
For the purposes of this paper, the difference between these two definitions is not important.
The other parameters that can  determine the effects of eclipsing are the stellar inclination angle, the stellar rotation period, the flare's duration and peak emission measure.

Under the assumption that the plasma density is uniform over the length of the flaring loop, the visible emission measure, $EM_{vis}(t)$, can be expressed as

\begin{equation} \label{eqn:EM}
EM_{vis}(t) = EM_{tot}(t) \left( \frac{V_{vis}(t)}{V_{tot}} \right)
\end{equation}

\noindent
where $EM_{tot}(t)$ is the flare emission measure curve that will be seen if the entire flaring loop is visible throughout the duration of the flare and $V_{vis}(t)/V_{tot}$ is the fraction of the volume of the flaring plasma that is visible at any given time $t$.
The quantity $V_{vis}(t)/V_{tot}$ is calculated at each time $t$ by considering a series of points equally spaced along the length of the flaring loop. 
Under the assumption that the loop thickness is small, the fraction of the flaring loop volume that is visible at this time is approximately equal to the fraction of these points that are visible.
We give details of how to determine whether a point on a flaring loop is eclipsed or visible in Appendix \ref{appendix:geocalculations}.
The methods used for chosing the function $EM_{tot}(t)$ is described in Paragraph 1 of Section \ref{sect:modelCOUP}.

The most obvious source of eclipsing of stellar flares comes from the host stars which we assume to be opaque spheres. 
However, other sources of eclipsing may be present.
In pre-main sequence stars, these may be binary companions, circumstellar discs, accretion columns extending from a circumstellar disc to the stellar surface, planets at small radii and flare associated prominences. 
In this paper, the only sources of eclipsing that we consider are host stars, circumstellar discs and flare associated prominences. 

We model circumstellar discs as opaque discs with smooth inner edges located at the equatorial corotation radii ($R_{co}=\left( G M_\ast/\omega^2\right)^{1/3}$, where $\omega$ is the angular velocity of the stellar surface at the equator) of their host stars.
The modelled discs are assumed to be flat and to lie in the equatorial plane.
The possibility of more complex discs is not considered here although it should be noted that a warped circumstellar disc could have a significant effect on a flare's lightcurve, particularly if the stellar inclination is such that a warped inner disc periodically obscures the view to the star (e.g. \citealt{2010A&A...519A..88A}).

We model prominences as opaque spheres that sit above the apex of flaring loops.
Thus, a prominence is characterised by its height above the flaring loop and its radius.
We take all prominences to be spheres of radius 0.5$R_\ast$, the centres of which have heights above the apex of the flaring loops of 0.55$R_\ast$. The prominences thus cover 25\% of the stellar disc, which is similar to the estimated projected areas of prominences on AB Dor and Speedy Mic (\citealt{1990MNRAS.247..415C,2006MNRAS.373.1308D}).

\section{The number of eclipsed flares} \label{sect:COUPflares}

In this section, we analyse the entire COUP sample in order to determine how many flares have been eclipsed.
We define non-eclipse candidate flares as flares that show a single rise followed by a single decay in their emission measure vs time curves (this includes both the 'typical' and the 'slow-rise top-flat' flares defined by \citealt{2008ApJ...688..418G}).
We define eclipse candidate flares as those that display sudden short duration decreases followed by increases in their emission measures. 

An eclipse that has a duration comparable to the duration of the flare will generally only result in a less luminous flare without a noticeably atypical morphology or a flare that is not visible at all.
For this reason, we expect that such dips should be found predominantly on long duration flares and rapidly rotating stars.

In order to illustrate the effects that eclipsing can have on typical flares, we show in Fig. \ref{fig:flarefits}, three examples of eclipse candidate flares from the COUP sample.
We demonstrate that an eclipse candidate flare can be produced by the eclipsing of a typical flare (i.e. flares consisting of a single rise phase followed by a slow exponential decay) by fitting model eclipsed flares to these observed emission measure vs time curves (for details, see the caption of Fig. \ref{fig:flarefits}).
We note, however, that while the forward problem of varying the model flare parameters to fit the observations is quite straightforward, the inverse problem of recovering the true flare parameters solely from the observations is in general not possible.

Of the 216 COUP flares catalogued by \citet{2008ApJ...688..418G}, we identify 62 (29\%) eclipse candidates. 
This number, given different levels of scepticism by the examiner, may be between 31 (14\%) and 71 (33\%). 
For the rest of this paper, we will take the value of 62 as the number of eclipse candidate flares in the COUP sample. 

Table \ref{tbl:COUPdist1} gives average values for several COUP flare and host star parameters, derived from parameters given by \citet{2008ApJ...688..418G}, for eclipse candidate and non-eclipse candidate flares separately. 
We have attempted to estimate by eye any decreases in the flare durations derived from the visible emission measure curves that might have been caused by eclipsing. 
This is only possible for flare emission measure curves which have been broken into two parts by large temporary eclipses.
It can be seen that contrary to expectations, the average flare durations as a fraction of host star rotation period is shorter for eclipse candidate flares when the original flare durations from \citet{2008ApJ...688..418G} are used.
When the durations are calculated assuming eclipsing has occurred, this is no longer a problem because the durations are always significantly longer than their original values. 
It can also be seen that the average peak emission measures are lower for eclipse candidate flares than for non-eclipse candidate flares which is consistent with eclipsing hypothesis.

\begin{figure}
\centering
\includegraphics[width=0.51\textwidth]{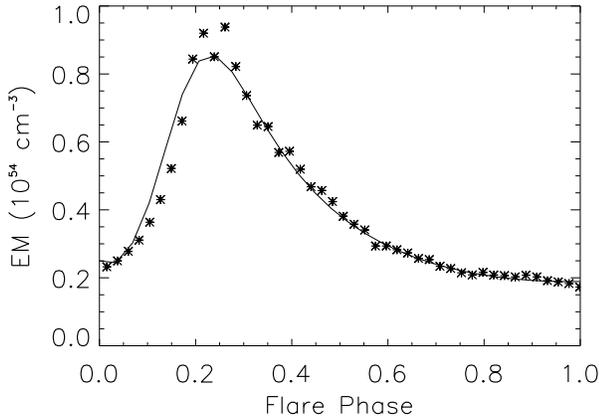}
\caption{Flare on COUP source 871 which our standard `typical' emission measure curve is based on. The full line represents the best fit to this curve which is used as the standard flare emission measure vs time curve.}
\label{fig:typical}
\end{figure}

\begin{figure}
\centering
\includegraphics[width=0.30\textwidth]{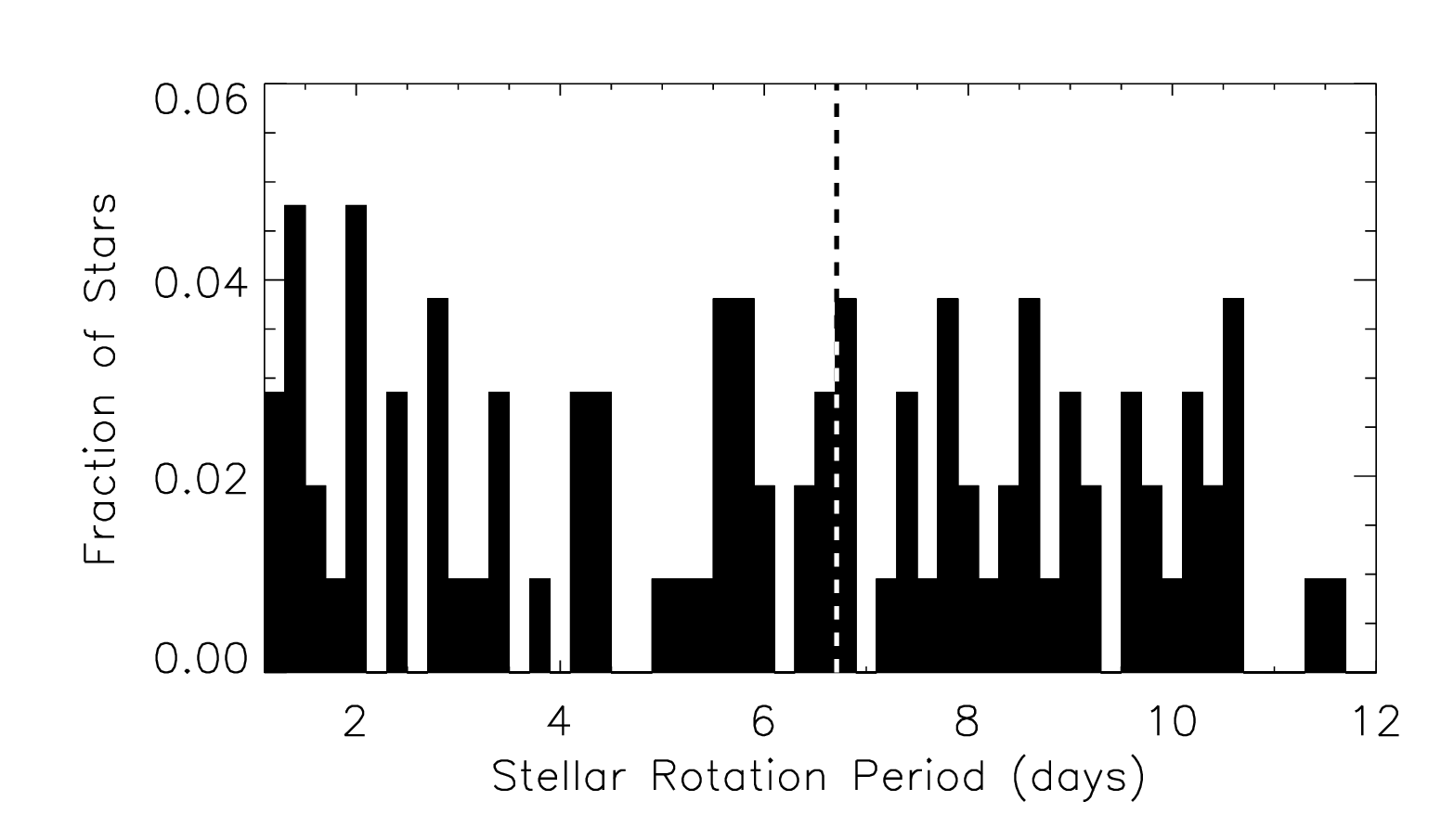}
\includegraphics[width=0.30\textwidth]{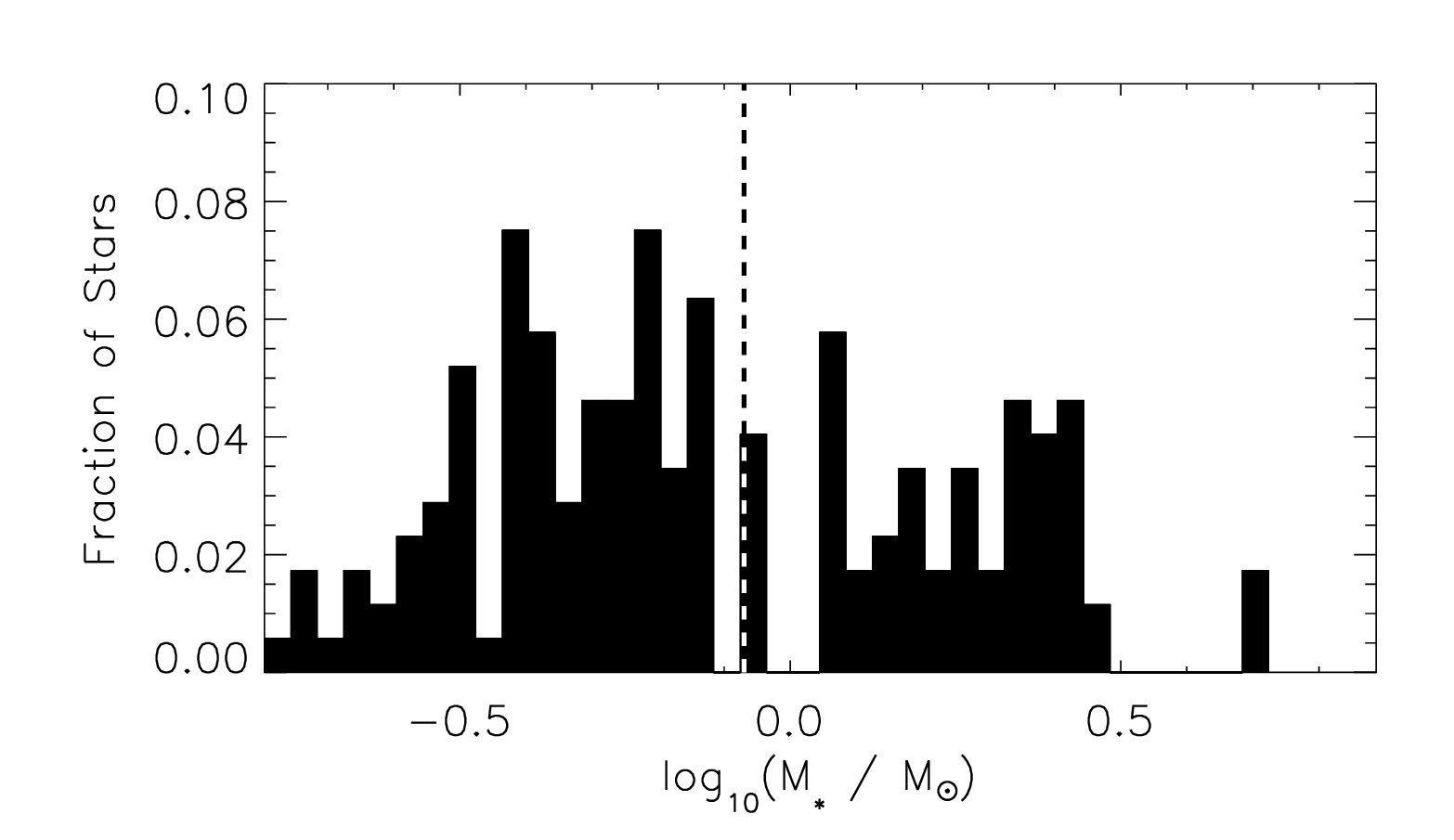}
\includegraphics[width=0.30\textwidth]{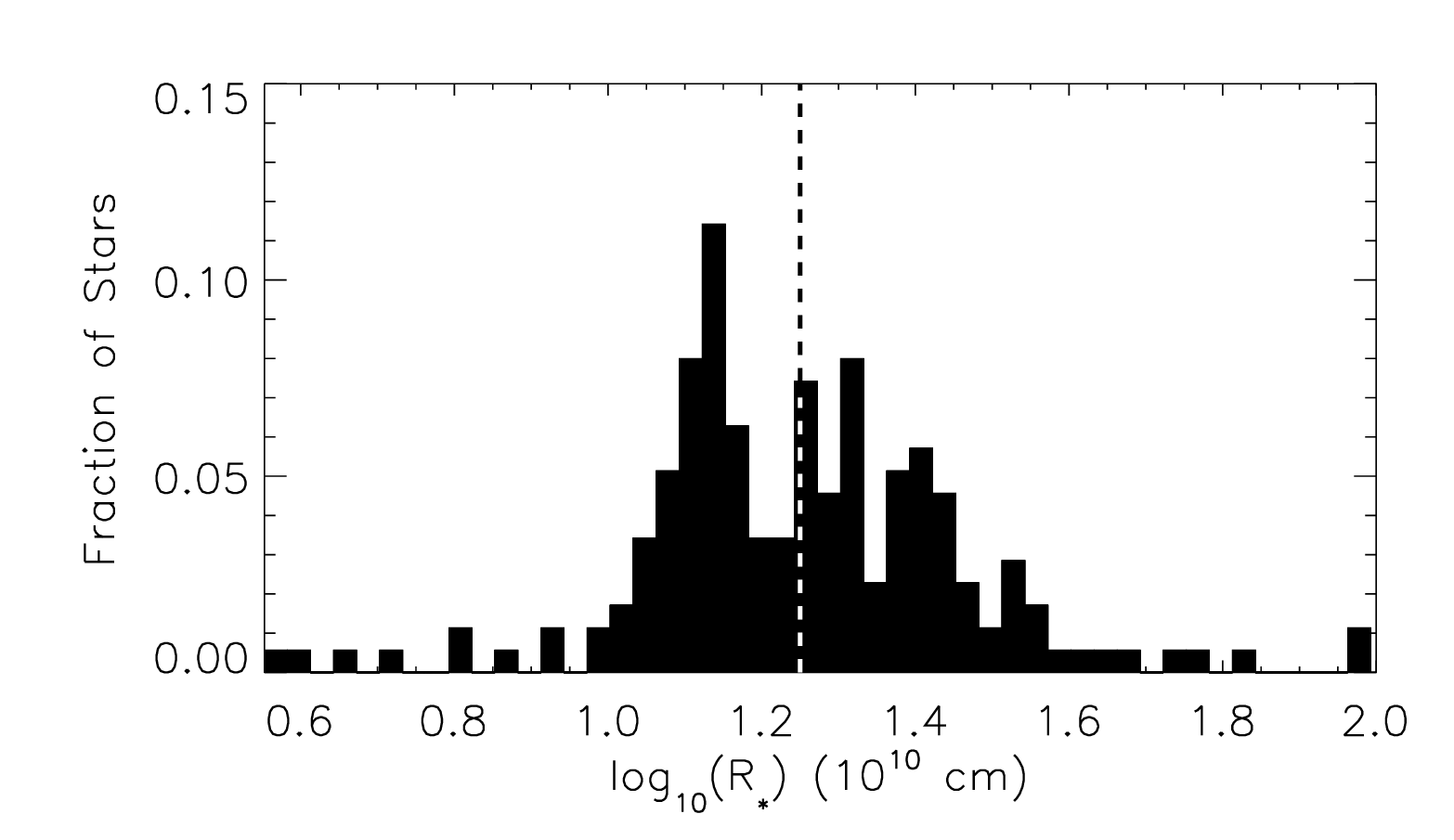}
\includegraphics[width=0.30\textwidth]{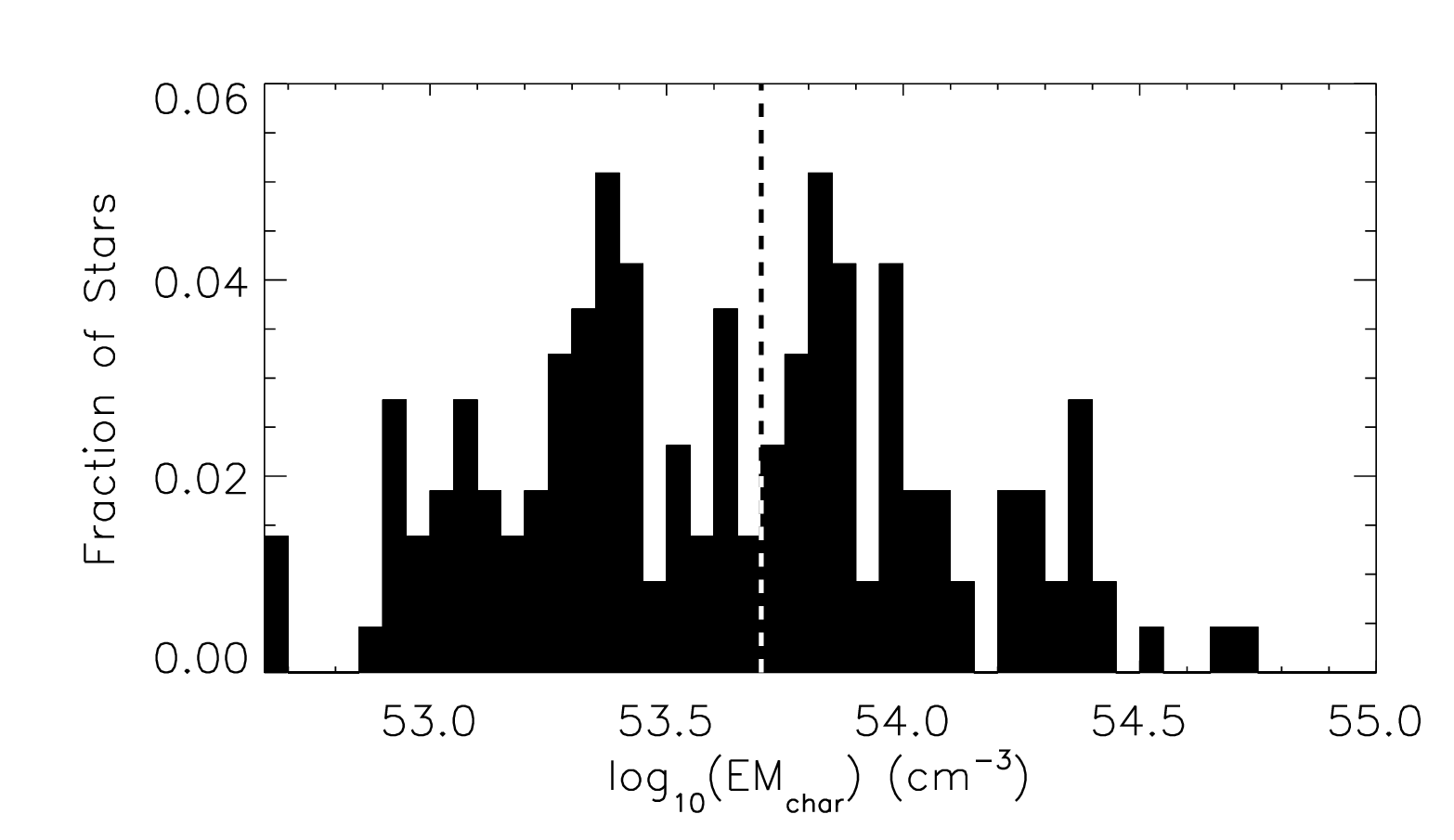}
\includegraphics[width=0.30\textwidth]{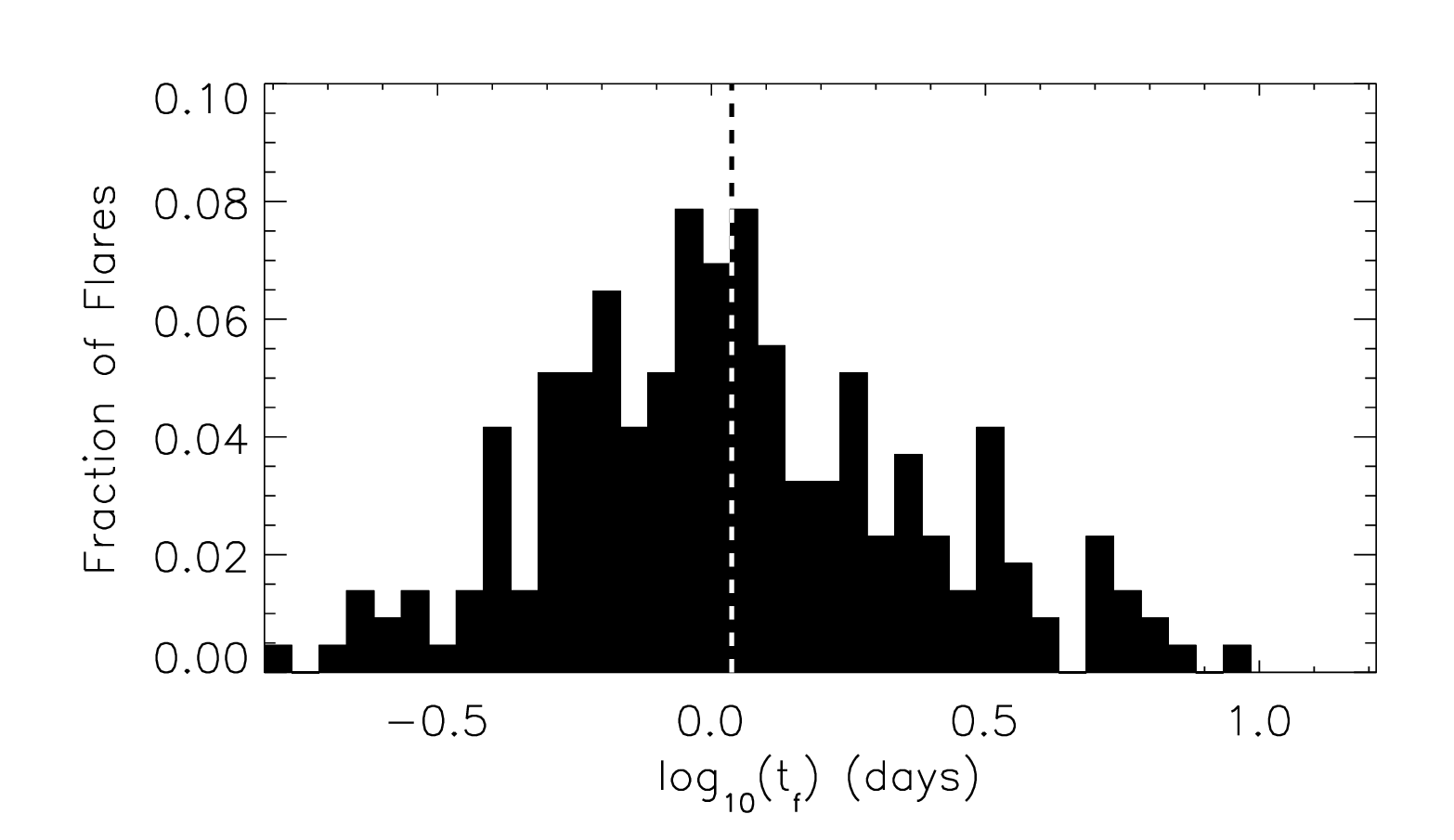}
\includegraphics[width=0.30\textwidth]{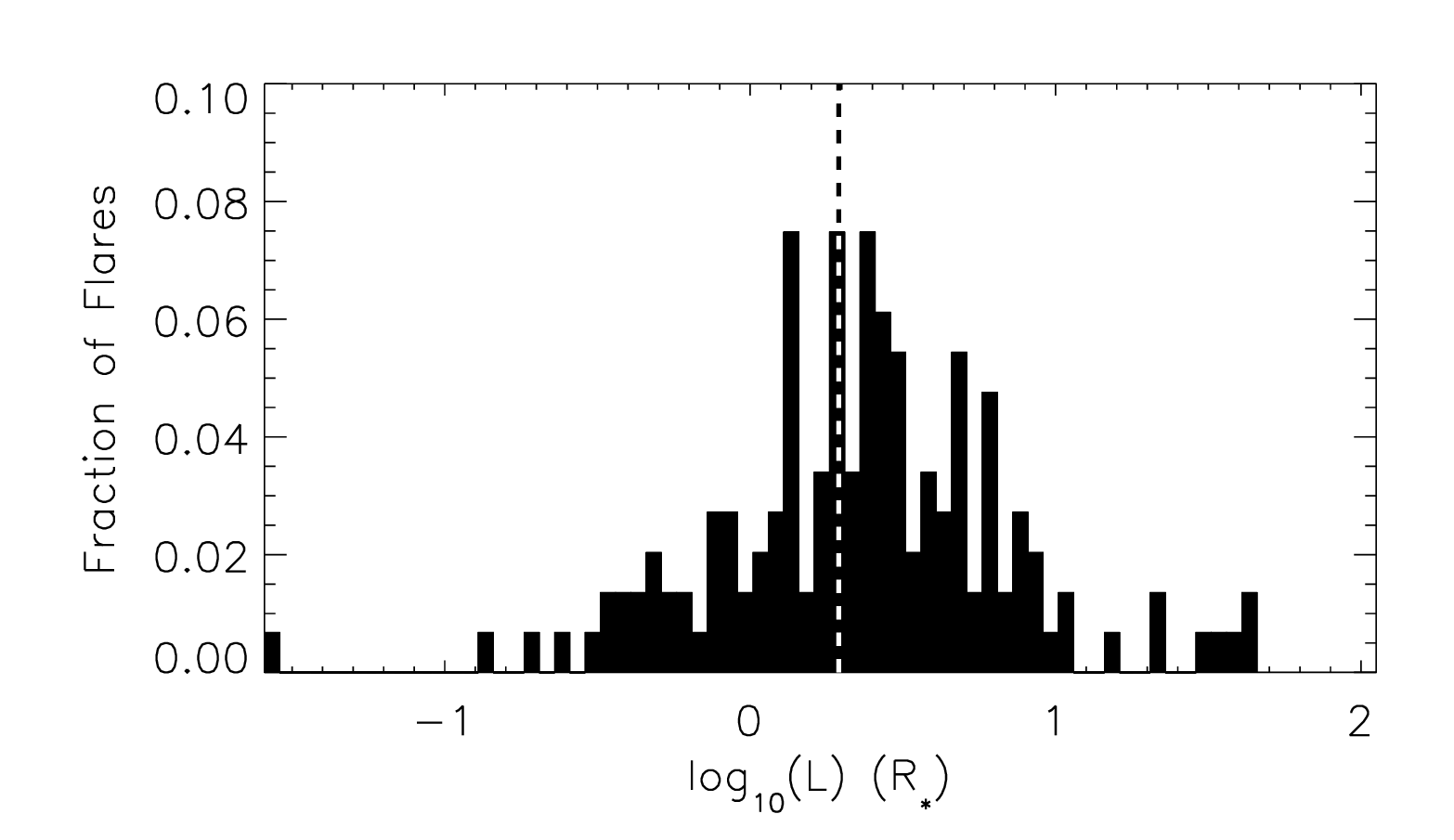}
\includegraphics[width=0.30\textwidth]{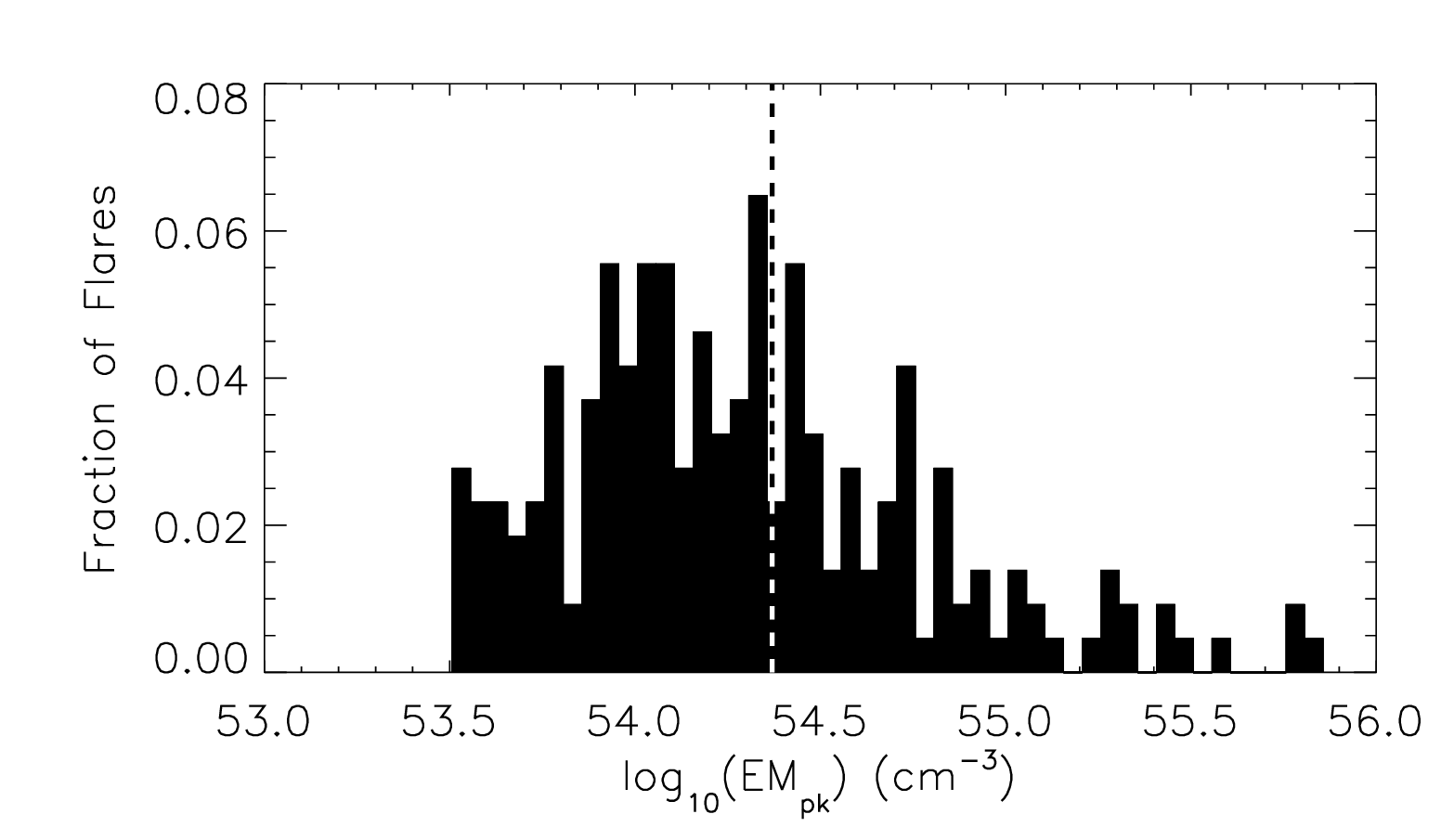}
\caption{Histograms showing the distributions of relevant stellar and flare parameters derived from the data given in Tables 1-3 of \citet{2008ApJ...688..418G}. The histograms show, from top to bottom, stellar rotation periods, stellar masses, stellar radii, host star characteristic emission measures, flare durations, flare loop lengths and flare peak emission measures. Dashed lines show the locations of the mean values.}
\label{fig:COUPparameterdistributions}
\end{figure}

\subsection{Modelling the set of COUP flares} \label{sect:modelCOUP}

Given that we classified 62 of the COUP flares as eclipse candidates, we now use our simple flare model to calculate the number of eclipse candidates that we would expect to see in the COUP flare sample. 
We assume that all flares are produced by the same energy release process acting within our simplified loop geometry, and therefore have similar typical intrinsic emission measure vs time curves (i.e. $EM_{tot}(t)$ in Eqn. \ref{eqn:EM}). 
We therefore choose the flare observed on COUP source 871 from the COUP sample (shown in Fig. \ref{fig:typical}) as the intrinsic flare profile. 
We then use this profile to produce many simulated flares by scaling the peak emission measure and the flare duration in the following way

\begin{equation}
EM_{tot,sim}(t)= EM_{tot,871} \left(t_{scaled} \right) \times \left(\frac{EM_{max,sim}}{EM_{max,871}} \right)
\end{equation}

\noindent where

\begin{equation}
t_{scaled} = t \times \frac{t_{sim}}{t_{871}}
\end{equation}

\noindent where the time $t$ is zero at the beginning of the impulsive phase of the flare, $EM_{tot,871}$, $EM_{max,871}$ and $t_{871}$ are respectively the intrinsic emission measure vs time curve, the maximum emission measure, and the duration for our standard flare and $EM_{tot,sim}$, $EM_{max,sim}$ and $t_{sim}$ are similar quantities for the simulated flare. 
We choose the values of $EM_{max,sim}$ and $t_{sim}$ randomly using the model described below in Section \ref{sect:probdist}. 
We note that although here we have focused on the emission measure vs time curve of the flare observed on COUP source 871 as our standard typical flare, repetitions of our analysis as discussed below using a different typical COUP flares yielded no significant difference in our results.

In order to determine the visible emission measures curves ($EM_{vis}(t)$) for the simulated flares we also need to specify the nine parameters that determine the loop geometry and position and the stellar rotation rate. 
Each of the randomly chosen parameters discussed above is chosen based on probability distributions that best approximate the distributions of these parameters in the COUP sample.
Where it is not possible to use observed distributions of parameters, reasonable assumptions, discussed below, must be made.
With these parameters we calculate the fraction of the flaring loops that are visible as a function of time (i.e. the fraction $V_{vis}/V_{tot}$ from Eqn. \ref{eqn:EM}) and using Eqn. \ref{eqn:EM}, we calculate the visible emission measure curves for each flare.

It is important in these calculations to define flares in the same way as \citet{2008ApJ...688..418G} in order that the results can be reliably compared.
For this reason, we define a flare as any energetic event with a peak emission measure exceeding four times the characteristic emission measure of the host star. 
In this way, flares that have been eclipsed to a an extent that they would not have been classified as bright flares in the COUP sample are discarded.

\subsection{Probability distributions for flare parameters} \label{sect:probdist}

In Section 2, we listed the nine geometric and temporal parameters that can affect, through eclipsing, the soft X-ray lightcurve morphologies of stellar flares. 
Another factor considered in this paper is the existence of other opaque material that can act as alternative sources of eclipsing. 
The two other sources of eclipsing considered here are circumstellar discs and flare associated prominences.
In order to calculate the radii of the inner edges of the circumstellar discs, which we assume to be at the equatorial corotation radius, we must also model the stellar masses and radii.
Thus, for the purposes of this paper, we must model eleven probability distributions.

Fig. \ref{fig:COUPparameterdistributions} shows histograms for seven of the parameters derived using the data given by \citet{2008ApJ...688..418G}. 
For these distributions, we ignore data from eclipse candidate flares.
With the exception of the stellar rotation periods, all these parameters can be modelled using log-normal distributions parametrised by their mean, $\mu$, and variance, $\sigma^2$.
The means and variances for these six parameters are given in Table \ref{tbl:COUPmoments}.
We assume that the stellar rotation periods have values that are evenly distributed between 0.1 and 11 days. 

The other parameters that need to be estimated in order to model the ONC flares are the starting longitudes of the flaring loops, $\phi_0$, the orientation of the loops, $\gamma$, the inclination angles of the stellar rotation axes to an observer's line-of-sight, $\theta_{view}$, and the colatitudes of the centres of the flaring loops, $\theta_{loop}$. 
The former two are taken to have values that are evenly distributed over all possible values. 
The latter two are taken to have a higher probability for values near the equator based on the probability density function $pdf(\theta)=\frac{1}{2} \cos \theta_{lat}$, where $\theta_{lat}$ is the latitude.

In these calculations, the presence of circumstellar discs around some of the stars is also considered.
\citet{2008ApJ...688..418G} derived near infrared colour excess ($\Delta(H-K_s)$) values for 140 of the flare host stars and used the condition $\Delta(H-K_s) > -0.06$ mag as a good indicator for the presence of circumstellar discs.
Of these 140 stars, 53 indicate the presence of a circumstellar disc. 
Thus, in the flares sets considered in the next section, where circumstellar discs are considered, each flare has a probability of 0.38 of having occurred on a star that has a disc.

It is also necessary to calculate the heights ($H$) and widths ($W$) of flaring loops when the only available information are the loop lengths.
For this reason, it is then necessary to assume a plausible relation between the heights and the widths of flaring loops.
We assume the relation that would be expected for a potential arcade with a maximum width of $W_{max}$.
This relation can be found in \citet{1986SoPh..106..335B} and is given by

\begin{equation} \label{eqn:HW}
\exp\left(-\frac{H}{R_\ast}\right) = cos\left(\frac{\pi W}{2 W_{max}}\right)
\end{equation}

\noindent
where $W_{max}$ is taken to be equal to $0.9R_\ast$ (if the value for $W_{max}$ is larger than $R_\ast$ the largest flare loops would not touch the stellar surface).
It is important to point out that even though we use the height-width relation for magnetic loops in a potential arcade, throughout this paper, the actual loop geometries are ellipses.

\begin{table*}
\begin{tabular}{cccccccccc}
\hline
 & $N_{tot}$ & $P_{rot}$ &  $EM_{pk}$ & $L$ & $N_{disc} / N_{tot}$ & \multicolumn{2}{c}{$t_f$ (COUP)} &  \multicolumn{2}{c}{$t_f$ (Modified)} \\
 & & (days) & $(10^{53} cm^{-3})$ & $(10^{10} cm)$ & & (days) & ($t_f / P_{rot}$) & (days) & ($t_f / P_{rot}$)\\ 
  & (1) & (2) & (3) & (4) & (5) & (6) & (7) & (8) & (9) \\ 
\hline
Eclipse Candidates & 62 & 7.08 & 33.92 & 42.71 & 0.33 & 1.28 & 0.27 & 2.36 & 0.47\\
\hline
Non-Eclipse Candidates & 154 & 6.57 & 46.87 & 29.22 & 0.44 & 1.59 & 0.36 & 1.59 & 0.36\\
\hline
All & 216 & 6.71 & 43.15 & 33.09 & 0.40 & 1.50 & 0.34 & 1.85 & 0.41\\
\hline
\end{tabular}
\caption{\emph{Average} values of flare and stellar parameters in the COUP set reported by \citet{2008ApJ...688..418G}. The data are presented for the entire set (bottom row) and separately for the flares that have been classified as eclipse candidates (top row) and non-eclipse candidates (middle row). The columns correspond to: Col. (1): the number of flares in each category. Col. (2): the host star rotation periods. Col. (3): the peak emission measures. Col. (4): the derived loop lengths. Col. (5): the fraction of flares on stars which show evidence that they possess a circumstellar disc. Cols. (6,7): the visible flare durations reported by \citet{2008ApJ...688..418G} as an absolute value and as a fraction of their host star's rotation period. Cols. (8,9): the flare durations after we attempt to correct for decreases in their visible durations due to noticeable eclipses.}
\label{tbl:COUPdist1}
\end{table*}

\begin{table*}
\begin{tabular}{ccccccc}
\hline
 & $log_{10}(t_f) (days)$ & $log_{10}(L) (R_\ast)$ & $log_{10}(M_\ast/M_\odot)$ & $log_{10}(EM_{pk}) (cm^{-3})$ & $log_{10}(EM_{char}) (cm^{-3})$ & $log_{10}(R_\ast) (10^{10}cm)$\\
  & (1) & (2) & (3) & (4) & (5) & (6)\\ 
\hline
$\mu$ & 0.037 & 0.30 & -0.07 & 54.37 & 53.71 & 1.25\\
$\sigma^2$ & 0.13 & 0.24 & 0.13 & 0.26 & 0.25 & 0.05\\
\hline
\end{tabular}
\caption{Mean and variance values for the six flare and host star parameters that are to be approximated as log-normal distributions. The columns correspond to: Col. (1): flare durations. Col. (2): Flare loop lengths. Col. (3): Mass of host stars. Col. (4): Flare peak emission measures. Col. (5): Host star characteristic emission measures. Col. (6): Host star radii. The values are calculated using data given by \citet{2008ApJ...688..418G}, which are plotted as histograms in Fig. \ref{fig:COUPparameterdistributions}.}
\label{tbl:COUPmoments}
\end{table*}

\begin{figure*}
\centering
\subfigure[]{\includegraphics[width=0.32\textwidth]{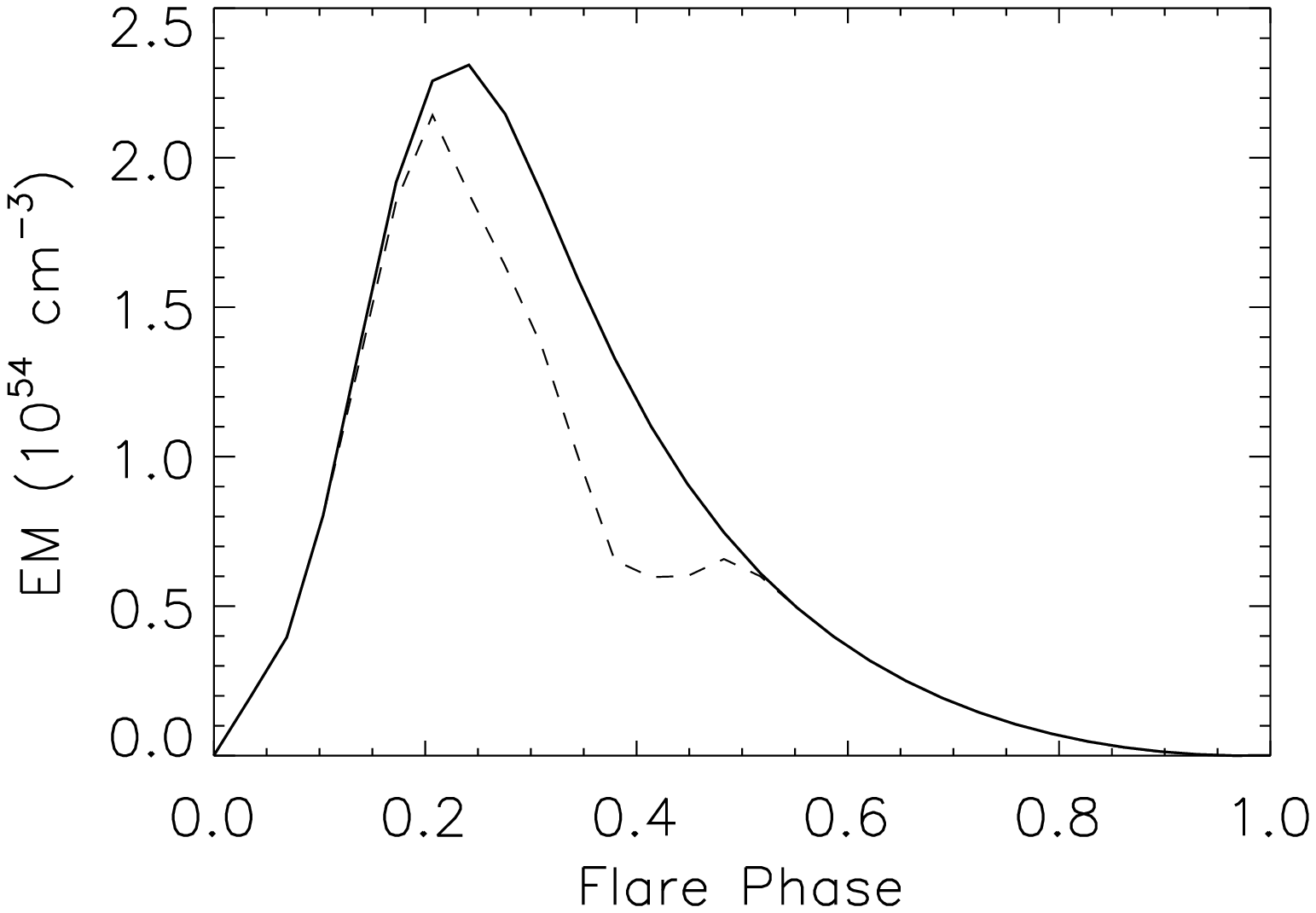}}
\subfigure[]{\includegraphics[width=0.32\textwidth]{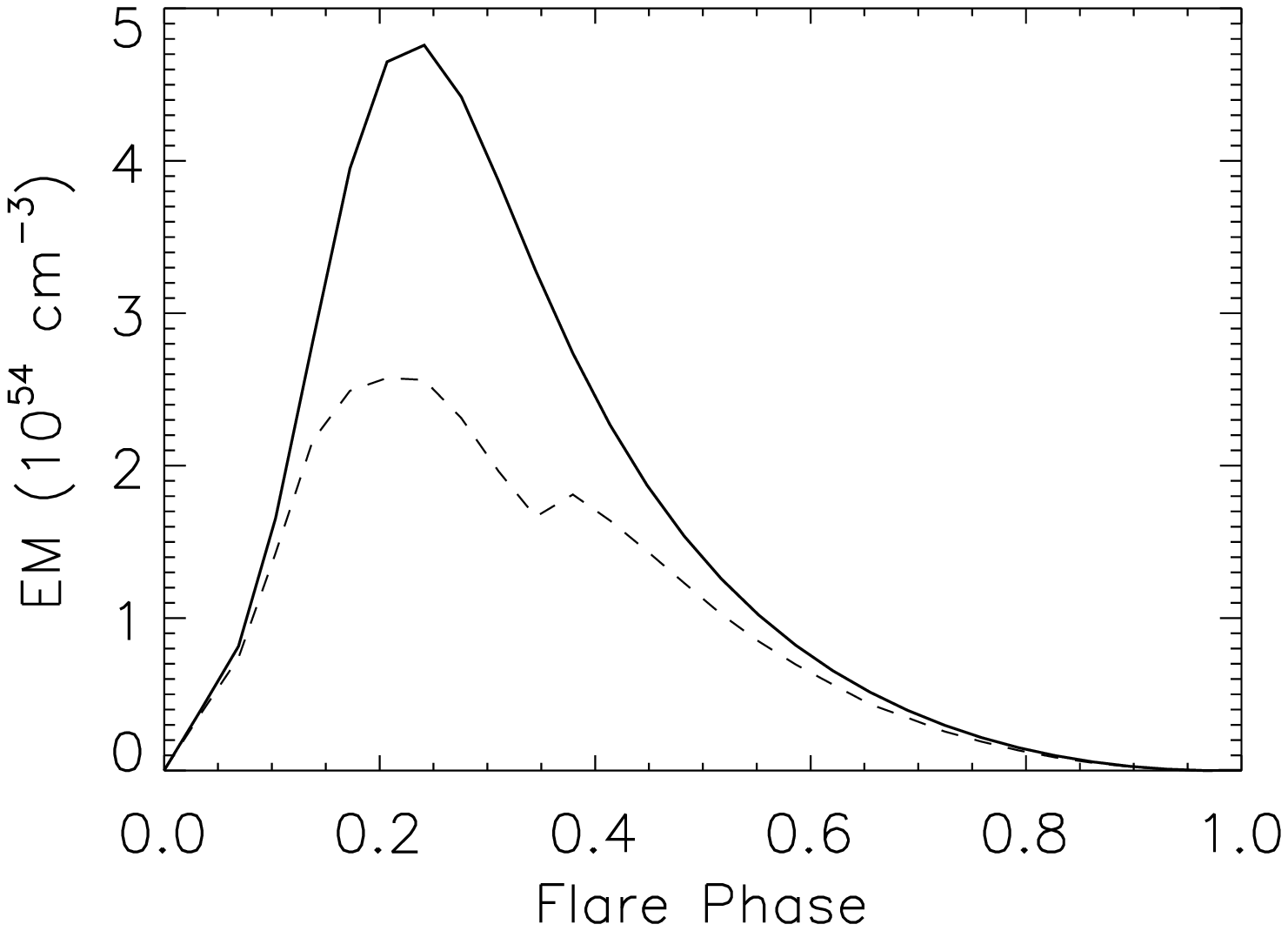}}
\subfigure[]{\includegraphics[width=0.32\textwidth]{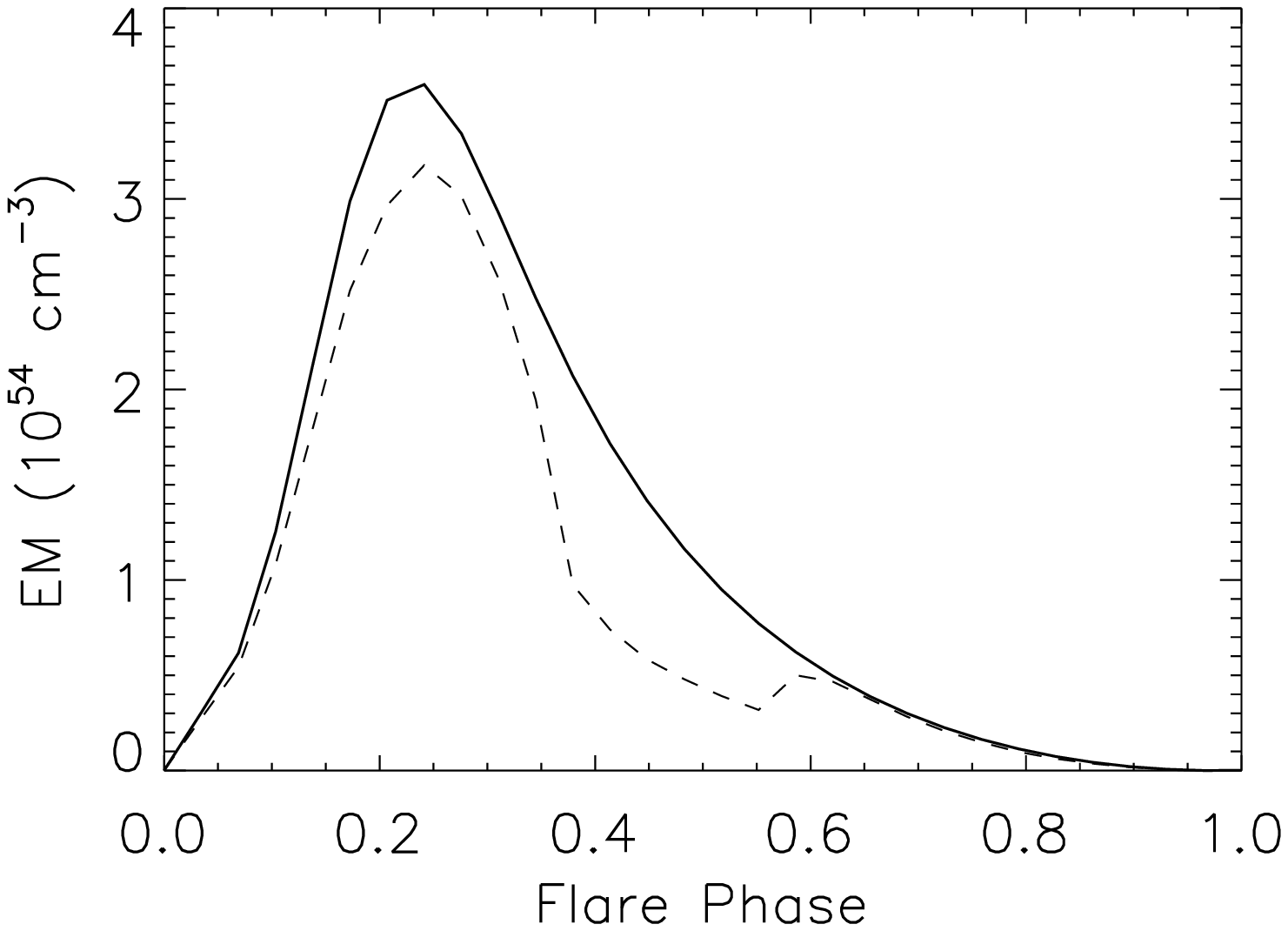}}
\subfigure[]{\includegraphics[width=0.32\textwidth]{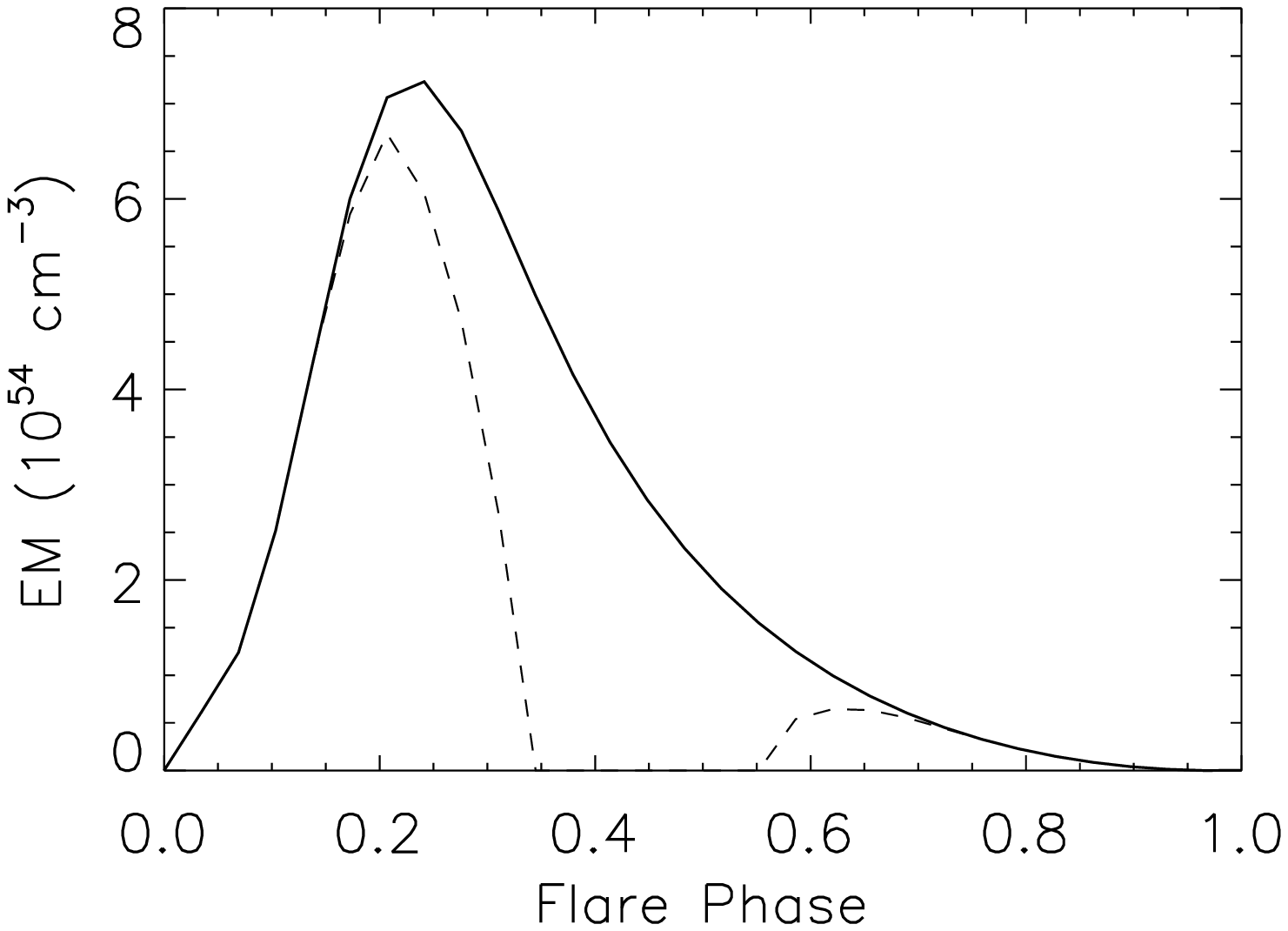}}
\subfigure[]{\includegraphics[width=0.32\textwidth]{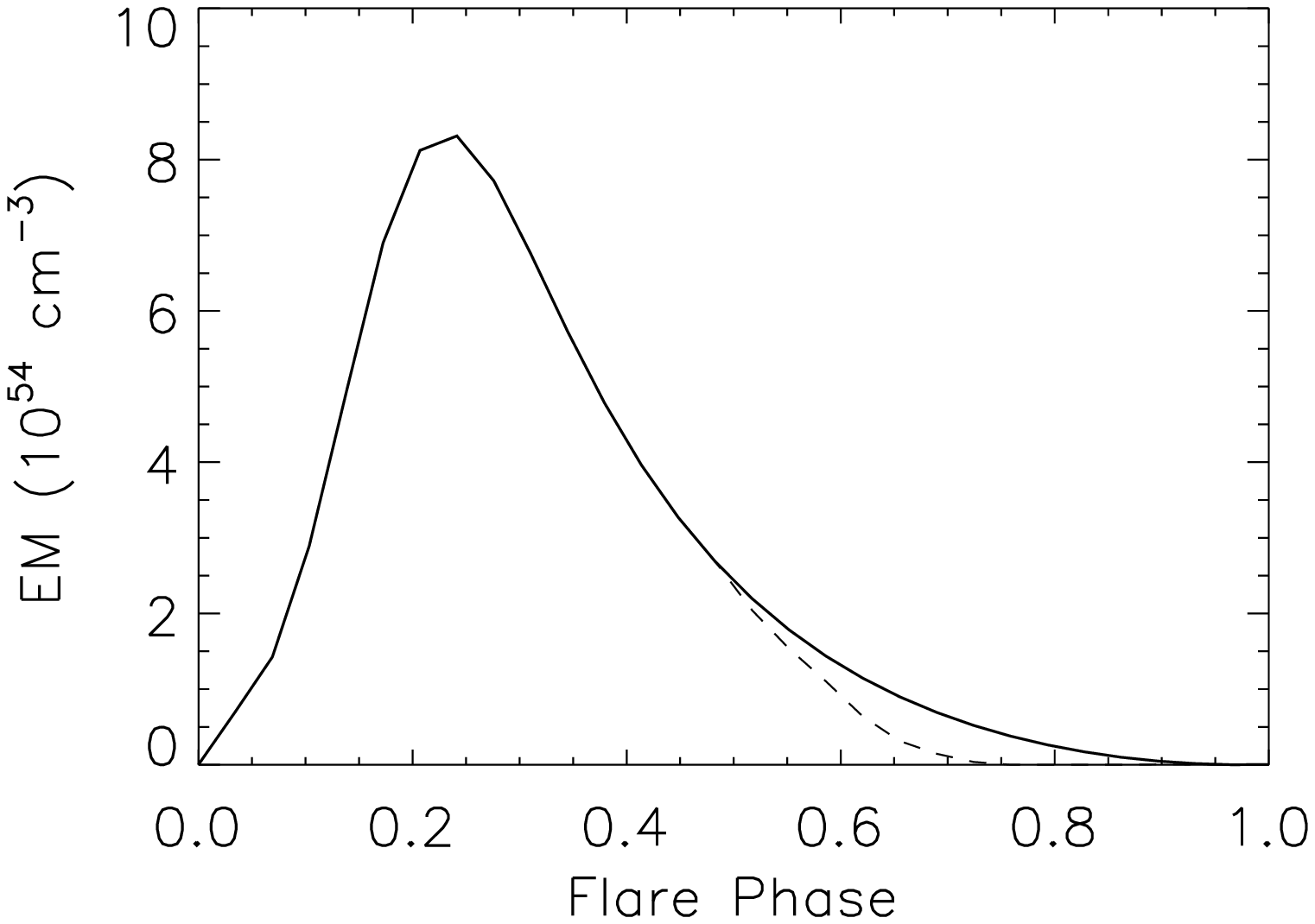}}
\subfigure[]{\includegraphics[width=0.32\textwidth]{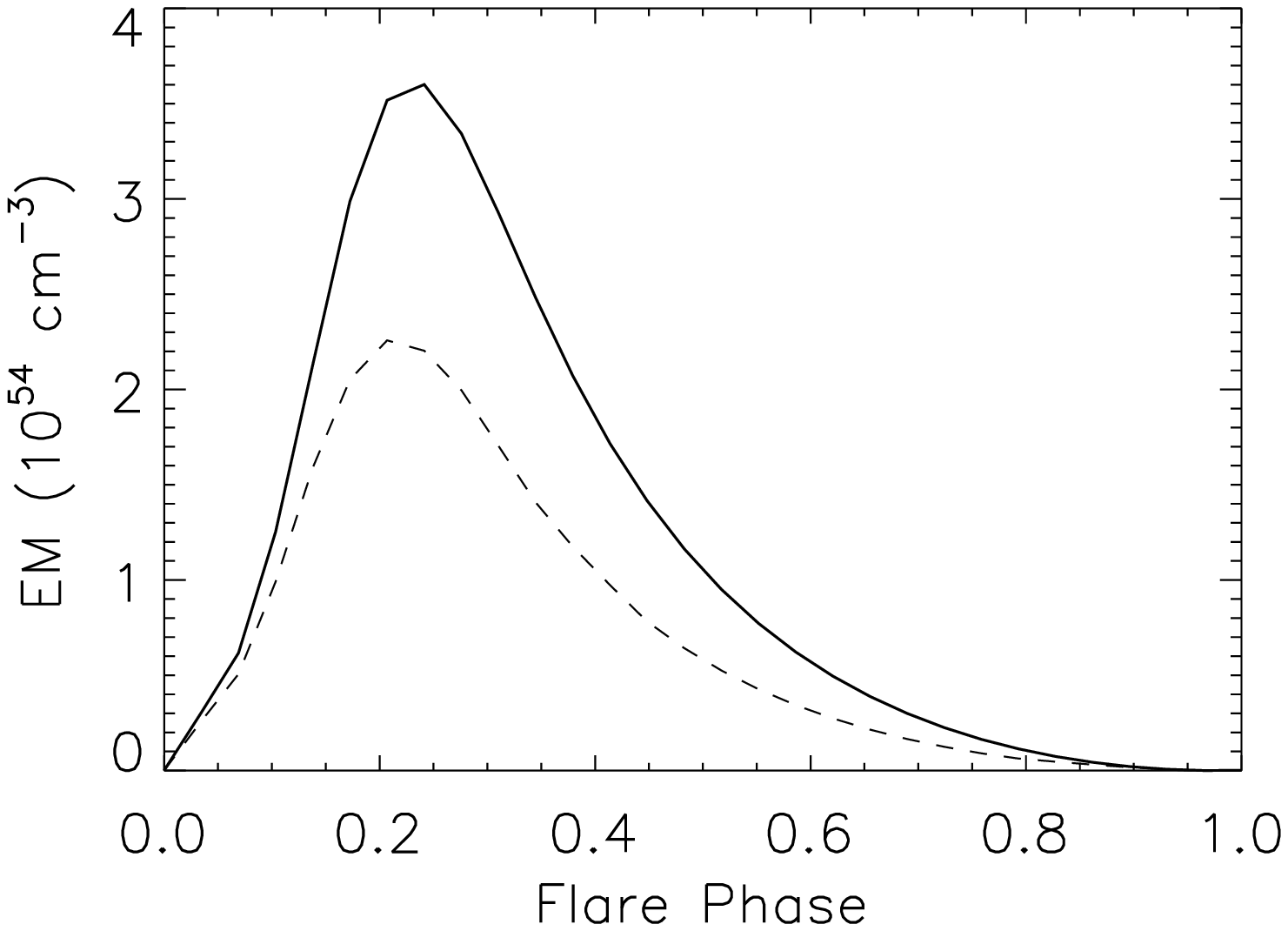}}
\caption{Six examples of modelled flares produced through eclipsing of the flaring plasma where the solid lines show the flare emission measure curves prior to eclipsing (i.e. the emission measure curve that would have been observed had the event always remained in view) and the dashed lines show the visible (i.e. eclipsed) emission measure curves. The latter arises by allowing the flare to enter or exit from rotational eclipse. These examples show that it is likely that atypical flare morphologies will be produced at random given a large sample of flares.}
 \label{fig:examplemodeled}
\end{figure*}

\begin{table*}
\begin{tabular}{ccccccccc}
\hline
 & $N_{tot}$ & $P_{rot}$  &  $EM_{pk}$ (visible) & $EM_{pk}$ (original) & \multicolumn{2}{c}{$t_f$ (visible)} & \multicolumn{2}{c}{$t_f$ (original)} \\
 & & (days) & $(10^{53} cm^{-3})$ & $(10^{53} cm^{-3})$ & (days) & ($t_f / P_{rot}$) & (days) & ($t_f / P_{rot}$) \\
  & (1) & (2) & (3) & (4) & (5) & (6) & (7) & (8)\\ 
\hline
Eclipse Candidates & 698 & 2.07 & 49.0 & 59.7 & 1.51 & 1.91 & 2.26 & 2.63\\
\hline
Non-Eclipse Candidates & 9302 & 5.77 & 46.5 & 49.6 & 1.31 & 0.47 & 1.33 & 0.48\\
\hline
All & 10,000 & 5.55 & 46.6 & 50.1 & 1.32 & 0.57 & 1.39 & 0.61\\
\hline
\end{tabular}
\caption{\emph{Average} values of flare and stellar parameters in the set of modelled flares. The data are presented for the entire set (bottom row) and separately for eclipse candidates (top row) and non-eclipse candidates (middle row). The columns correspond to: Col. (1): the number of flares in each category. Col. (2): the stellar rotation periods. Col. (3): the visible peak emission measures. Col. (4): the \emph{actual} peak emission measure. Cols. (5,6): the visible flare durations as an absolute value and as a fraction of the host stars' rotation periods. Cols. (7,8): the \emph{actual} flare durations \emph{prior to eclipsing}.}
\label{tbl:modelCOUPdist1}
\end{table*}

\subsection{Results} \label{sect:results}

Using the method described above, we model a set of 10,000 flares. 
In this section, we analyse these flares using the same method of visual inspection to select eclipse candidate flares as was used to analyse the COUP sample. 
In the initial results presented here, circumstellar discs and flare-associated prominences are not considered.
These are included in separate results presented at the end of this section.

Due to eclipsing, not every modelled flare has a peak emission measure that is large enough to be classified as a bright flare in the COUP observations. 
In order to obtain 10,000 flares which make it through the observational selection criteria adopted by \citet{2008ApJ...688..418G} it is necessary to model 10,878 flares in total. 
By inspecting the entire sample of modelled flares, we classify only 7.0\% as eclipse candidates.
This small value is to be contrasted with the larger number of 29\% eclipse candidate flares in the COUP sample.

Six examples of flare emission measure curves which have been affected by eclipsing are shown in Fig. \ref{fig:examplemodeled}.
We determine that eclipse candidates are likely to occur in a sample of 216 flares under the conditions present in the ONC. 
Thus, we conclude that it is likely that a number (some, although not necessarily all) of the eclipse candidate flares in the COUP sample have been produced by the rotational eclipsing of typical flares.
However, it is unlikely that the entire sample has been produced in this way. 
Given that the probability of one flare being an eclipse candidate is 0.07, using the Binomial distribution, we estimate that the probability of 62 flares being eclipse candidates in a sample of 216 flares is approximately $10^{-25}$. 
In order to explain the large number of atypical COUP flares, it is thus necessary to assume other physical mechanisms, such as multiple heating events in a single flaring loop, multiple flares whose lightcurves have been superimposed, or stellar analogues of solar coronal arcades where a reconnection event triggers subsequent events and associated flares/heating of neighbouring loops.

\begin{figure}
\centering
\includegraphics[width=0.4\textwidth]{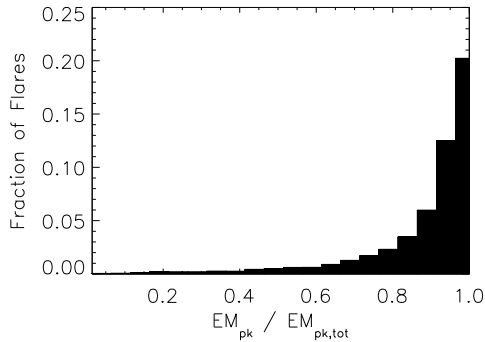}
\caption{Histogram showing the visible peak emission measures, $EM_{pk}$, as a fraction of their uneclipsed values, $EM_{pk,tot}$, for the set of modelled flares. Only the 105 flares which had their peak emission measures decreased by eclipsing are included here.}
 \label{fig:peakemdrop}
\end{figure}

The eclipse candidates in the modelled set of flares, however, do not represent the full sample that have undergone eclipsing.
The fifth and sixth flares shown in Fig. \ref{fig:examplemodeled} show examples of flares that have been eclipsed but still have `typical' emission measure curves.  
A total of 63\% of the modelled flares have been partially eclipsed for at least a fraction of their durations and 49\% of the flares have their peak emission measures reduced.
However, in most cases, eclipsing leads to an insignificant decrease in the visible emission measure value.
This can be seen in Fig. \ref{fig:peakemdrop}.
Of the modelled flares, 6\% show significant decreases in their visible durations.
Therefore, we expect that analysis of such flares may lead to derived flare parameters that are different from the true physical properties of the magnetic structure containing the flare; we explore this point further in the next section.

One type of flare morphology seen in the COUP sample but not in the flares modelled here are the slow-rise top-flat flares, defined by \citet{2008ApJ...688..418G}.
Such morphologies can be produced through eclipsing, as can be seen in the second peak in the fourth example shown in Fig. \ref{fig:examplemodeled}.
However, this peak has not made it through the selection criterion for flares so it is not counted in the sample of modelled flares. 

In Table \ref{tbl:modelCOUPdist1}, we give average values for flare and host star parameters, both before and after eclipsing has been taken into account, for eclipse candidate and non-eclipse candidate flares separately.
It can be seen that eclipsing causes a reduction in the average durations and peak emission measures.
It can also be seen that longer duration flares on faster rotating stars are more likely to be eclipse candidates, as expected for flares randomly distributed in latitude and longitude.

One reason why there may be more eclipse candidates seen in the observed COUP sample than in the modelled flare set could be that the COUP flares are being eclipsed by circumstellar discs or flare-associated prominences.
To investigate to what extent this may be the case, we repeat the above calculations to produce two more sets of 10,000 flares.
In the first set, we assume that circumstellar discs are present around 38\% of the host stars; the same fraction as determined \citet{2008ApJ...688..437G} based on Spitzer H-K excess emission.
With this assumption, we find that 6.4\% are eclipse candidates. 
In order to produce 10,000 visible flares, it was necessary to produce 11,689 flares in total.
In the second set, we assume that prominences are present above the apex of each flaring loop.
With this assumption, we find that 7.7\% are eclipse candidates.
In order to produce 10,000 visible flares, it was necessary to produce 11,612 flares in total.
Thus, it is clear that even with circumstellar discs and flare-associated prominences we are not able to explain all of the eclipse candidates seen in the COUP sample with eclipsing of single magnetic loops only.

\section{Flare Loop Lengths} \label{sect:lengths}

A common method for determining the loop length of an unresolved stellar flare involves the comparison of the flare's emission measure and temperature data with hydrodynamic flare models. This method, detailed by \citet{1997A&A...325..782R}, defines the loop half-length as

\begin{equation}
L (cm) = \frac{\tau_d (ks) \sqrt{T_{pk} (MK)}}{3.7 \times 10^{-4} F(\zeta)}
\end{equation} 

where 

\begin{equation}
F(\zeta) = \frac{0.63}{\zeta-0.32}+1.41
\end{equation} 

\begin{equation}
T_{pk}=0.068 T_{obs}^{1.2}
\end{equation} 

\noindent where $\tau_d$(ks) is the time that it takes for the flare's emission measure to decay by a factor of $e$ (the e-folding timescale), $T_{pk}$(MK) is the temperature at the apex of the flaring loop when the flare's emission measure is at its peak, $T_{obs}$(MK) is the observed average loop temperature at this time and $\zeta$ is the gradient of the decay phase of the $log(EM^{\frac{1}{2}})$ - $log{T}$ plot. See \citet{1997A&A...325..782R} for full details (for a discussion on the validity of the single loop model, see Appendix A of \citet{2011ApJ...730....6G}). 

In order to explore the effect that eclipsing can have on the loop lengths derived from this method, which has been commonly employed in the analysis of flares on young stars (e.g. \citet{2005ApJS..160..469F}), we take the emission measure and temperature data for the `typical' flare seen on COUP source 871 and produce a large set of 10,000 flares using the method described in Section \ref{sect:probdist}.
In these calculations, however, we only pick at random the stellar parameters (e.g. radius, rotation period) and the locations and orientations of the flaring loops.
The parameters specific to the flare (e.g. peak emission measures, loop length, flare duration) are kept at the values derived by \citet{2008ApJ...688..418G}. 
We then apply the loop length analysis to the uneclipsed and eclipsed flare emission measure curves, assuming that eclipsing has no effect on the determined temperatures. 

\begin{figure}
\includegraphics[width=0.49\textwidth]{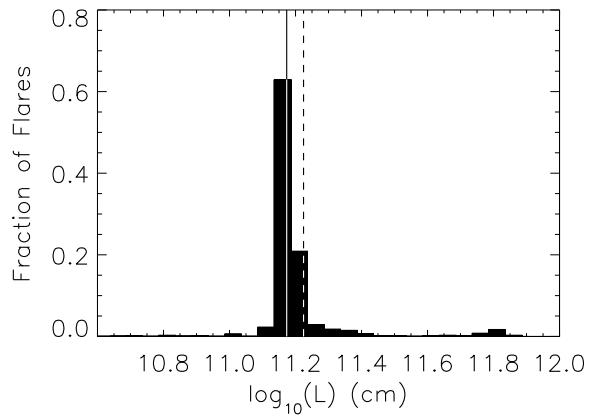}
\caption{Histogram showing the derived lengths of flaring loops after eclipsing has been taken into account (flares which have not undergone any eclipsing are included). The solid vertical line shows the length that would have been derived for all of the flares if no eclipsing had occurred. The dashed vertical line shows the log of the average derived loop lengths (not the average of the log).}
 \label{fig:lengths}
\end{figure}

The effect of eclipsing is shown in Fig. \ref{fig:lengths}.
The loop length based on the uneclipsed emission measure curve is $14.9 \times 10^{10}$cm. 
After eclipsing has been taken into account for a large set of flares, the average calculated loop length for that set is increased to $16.8 \times 10^{10}$cm. 
This simple estimate suggests that in most cases eclipsing has little effect on derived loop lengths.

\section{Summary and Conclusions} \label{sect:summary}

Although most stellar flares have typical soft X-ray lightcurve morphologies (i.e. a single rapid rise followed by a slow exponential decay), many flares have atypical morphologies. 
Many of these show multiple peaks or small dips in their lightcurves (\citealt{2008ApJ...688..418G}). 
Based on solar analogies, such flares are often interpreted as being the result of multiple heating events in the same flaring loop or the superposition of separate overlapping flares (for example, see \citealt{2004A&A...416..733R}; \citealt{2005A&A...430..155P}; \citealt{2008MNRAS.387.1627P}; \citealt{2010ApJ...712...78L}). 
In this paper, we have considered an alternative geometric interpretation in which these atypical flare morphologies are produced by the eclipsing of flaring plasma due to the rotation of the host star.
This interpretation has been considered for individual flares in previous studies (\citealt{1997ApJ...486..886S}, \citealt{1999A&A...344..154S}, \citealt{1999Natur.401...44S}, \citealt{2003A&A...412..849S}, \citealt{2006A&A...445..673S}, \citealt{2007A&A...466..309S}).

Using data from the \emph{Chandra} Orion Ultra-deep Project, \citet{2008ApJ...688..418G} identified 216 stellar flares on 161 PMS stars.
As the COUP sample contains a range of flare morphologies, in this paper, we have used it to explore the eclipsing interpretation.
We analysed the entire COUP sample by eye to determine which of them are eclipse candidate flares.
In Fig. \ref{fig:flarefits}, we took three examples of these and showed that their emission measure vs time curves can easily be produced by the eclipsing of typical flares.
In Section \ref{sect:COUPflares}, we showed that the entire COUP sample contained 62 (29\%) eclipse candidates.
However, by producing a large modelled set of flares similar to the COUP sample, we showed that although 63\% of the modelled flares underwent eclipsing, this was detectable in only 7.0\% of them. 
In Section \ref{sect:lengths}, we showed that eclipsing can effect the derived loop lengths for flares, but in most cases such an effect is negligible.

Our conclusions from these results are as follows

\begin{itemize}
\item The time variable eclipsing of stellar flares contained within single magnetic loop structures can produce the atypical morphologies observed in the COUP sample. 
Thus, given a flare with an atypical lightcurve, it is not necessary to invoke unusual physical mechanisms to account for the flare's morphology. 
However, it should be noted that eclipsing is much more likely to cause an atypical morphology on longer duration flares and on more rapidly rotating stars.
\item However, the observed frequency of eclipse candidate flares in the COUP sample is far higher than we would expect if eclipsing was the only mechanism by which atypical flares were being produced. 
Thus, alternative physical mechanisms, such as the stellar analogies of solar coronal arcades, must be responsible for most of the atypical COUP flares.
\item Even in cases where an observed flare has a typical morphology, it is not possible to know from the flare lightcurve alone whether or not eclipsing has taken place.
\item Eclipsing is unlikely to have a significant effect on derived loop lengths.
\end{itemize}

\section{Acknowledgments}

CJ acknowledges support from an STFC studentship. 
SGG acknowledges support of the Science and Technology Facilities Council (grant number ST/G006261/1).
KVG acknowledges support from the Chandra ACIS Team contract SV4-74018 (G. Garmire, PI), issued by the Chandra X-ray Center operated by the Smithsonian Astrophysical Observatory for and on behalf of NASA under contract NAS8-03060.

\appendix

\section{Conditions for Eclipsing} \label{appendix:geocalculations}

\begin{figure}
\centering
\includegraphics[width=0.5\textwidth]{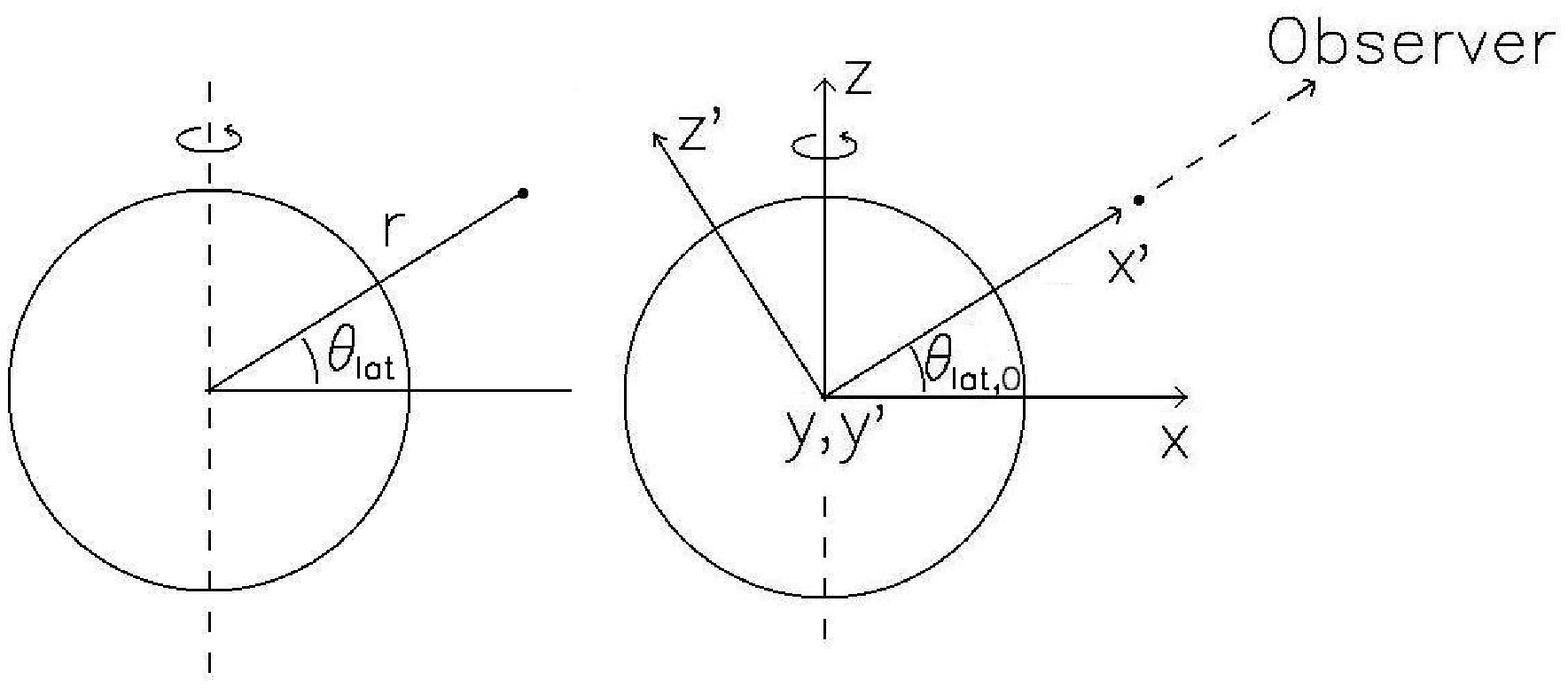}
\caption{The various coordinate systems used to determine whether or not an object has been eclipsed. In this image, the $y$ and $y'$ axes are going into the page.}
 \label{fig:coordinatesystems}
\end{figure}

In this appendix, we describe the method used in this paper to determine whether or not a point is being eclipsed by either the host star, a circumstellar disc or a prominence. 

It is easiest to set up and evolve a stellar system in a spherical polar coordinate system ($r, \theta_{lat}, \phi$).
However, the conditions for eclipsing are simplest when the system is represented in Cartesian coordinates with either one axis pointing along the line of zero longitude and latitude ($x,y,z$) or pointing towards the observer ($x',y',z'$). 
These three coordinate systems can be seen in Fig. \ref{fig:coordinatesystems}.
In all three coordinate systems, the origin is at the centre of the host star.
The transformations between these three coordinate systems are as follows. 
\begin{eqnarray}
x=r \cos\phi \cos\theta_{lat}\\
y=r \sin\phi \cos\theta_{lat}\\
z=r \sin\theta_{lat}\\
x'=z \sin\theta_{lat,0} + x \cos\theta_{lat,0}\\
y'=y\\
z'=z \cos\theta_{lat,0} - x \sin\theta_{lat,0} 
\end{eqnarray}

where $\theta_{lat}$ is the latitude and defined such that it has values between $-90^o$ and $90^o$ and the observer is located at $(r,\theta_{lat},\phi)$ equal to $(\infty,\theta_{lat,0},0)$ and $(x',y',z')$ equal to $(\infty,0,0)$. 

Consider a point which does not lie within the star, the disc, or the prominence, and has coordinates $(r,\theta_{lat},\phi)$, $(x,y,z)$ and $(x',y',z')$ in the different coordinate systems. 
This point is only visible if it is not eclipsed in all three of the following conditions.

\emph{Condition 1:} The point is eclipsed by the host star with radius $R_\ast$ if 

\begin{equation}
x' < 0
\end{equation}

\begin{equation}
y'^2 + z'^2 < R_\ast^2 
\end{equation}

The latter condition is only met if the point coincides with the disc of the host star in the plane of the sky and the former condition tests if the point is behind or in front of the host star.

\emph{Condition 2:} The point is eclipsed by a circumstellar disc with an inner hole with a radius of $R_{trunc}$, which we assume to be the equatorial corotation radius, if 

\begin{equation} \label{eqn:firstdisccondition}
\theta_{lat} \times \theta_{lat,0} < 0
\end{equation}

\begin{equation}
\sin^2{\theta_{lat,0}} z'^{2} + y'^{2} > R_{trunc}^2
\end{equation}

As $\theta_{lat}$ and $\theta_{lat,0}$ are defined such that they are positive in one hemisphere and negative the the other, the first of these condition is only met if the point and the observer are in opposite hemispheres of the star.
The second condition then tests if the point is visible through the inner hole of the disc as projected onto the plane of the sky.

\emph{Condition 3:} The point is eclipsed by a prominence of radius $R_p$ with its centre at $(r_p,\theta_{lat,p},\phi_p)$, $(x_p,y_p,z_p)$ and $(x'_p,y'_p,z'_p)$ if 

\begin{equation}
x' < x'_p
\end{equation}

\begin{equation}
(y'-y'_p)^2 +(z'-z'_p)^2 < R_p^2
\end{equation}

These conditions are the same as the conditions for eclipsing by the host star above with corrections for a sphere with a different radius and which is not centred at the origin of the coordinate system.

\bibliographystyle{mn2e}
\bibliography{mybib}

\end{document}